\documentclass[twocolumn, prd,reprint,nofootinbib,amsmath,amssymb,aps,floatfix]{revtex4-1}
\usepackage{amssymb,amsmath,graphicx}
\usepackage{color}
\usepackage[normalem]{ulem}
\usepackage{dsfont}
\usepackage{braket}

\newcommand{\be}{\begin{equation}}
\newcommand{\ee}{\end{equation}}

\begin{document}
\title{Ignorance is Cheap: From Black Hole Entropy To Energy-Minimizing States In QFT}
\author{Raphael Bousso}
\email{bousso@lbl.gov}
\author{Venkatesa Chandrasekaran}
\email{ven.chandrasekaran@berkeley.edu}
\author{Arvin Shahbazi-Moghaddam}
\email{arvinshm@gmail.com}
\affiliation{Center for Theoretical Physics and Department of Physics\\
University of California, Berkeley, CA 94720, USA 
}%
\affiliation{Lawrence Berkeley National Laboratory, Berkeley, CA 94720, USA}
\begin{abstract}

Behind certain marginally trapped surfaces one can construct a geometry containing an extremal surface of equal, but not larger area. This construction underlies the Engelhardt-Wall proposal for explaining the Bekenstein-Hawking entropy as a coarse-grained entropy. The construction can be proven to exist classically but fails if the Null Energy Condition is violated.

Here we extend the coarse-graining construction to semiclassical gravity. Its validity is conjectural, but we are able to extract an interesting nongravitational limit. Our proposal implies Wall's ant conjecture on the minimum energy of a completion of a quantum field theory state on a half-space. It further constrains the properties of the minimum energy state; for example, the minimum completion energy must be localized as a shock at the cut. We verify that the predicted properties hold in a recent explicit construction of Ceyhan and Faulkner, which proves our conjecture in the nongravitational limit.

\end{abstract}
\maketitle

\section{Introduction and Summary}
\label{sec-intro}

There is a remarkable interplay between testable low-energy properties of quantum field theory (QFT), and certain conjectures about quantum gravity, in which the area of surfaces is associated to an entropy. For example, the classical focussing theorem in General Relativity relies on the Null Energy Condition and so can fail in the presence of quantum matter. A Quantum Focussing Conjecture (QFC) was proposed to hold in the semiclassical regime; it implements a quantum correction to the classical statement by replacing the area with the area plus exterior entropy, i.e., the ``generalized entropy.'' This was a guess about quantum gravity, but it led to a new result in QFT. Namely, the Quantum Null Energy Condition (QNEC) was discovered as the QFT limit of the QFC~\cite{Bousso:2015mna}.

The QNEC has since been laboriously proven within relativistic quantum field theory~\cite{Bousso:2016aa, Balakrishnan:2017aa, Ceyhan:2018zfg}. The fact that the QNEC arises more directly and simply from a hypothesis about quantum gravity is striking. Experimental tests of the QNEC may be viable and should be regarded as test of this hypothesis.

Here we will discover a related but distinct connection of this type. We begin again with a classical gravity construction, though one motivated by quantum gravity. The notion that black holes carry Bekenstein-Hawking entropy (proportional to their area) has been fruitful and widely explored, but we stress here that it is a hypothesis that has not been experimentally tested. This hypothesis leads to a puzzle: if the black hole was formed from a pure state, then the entropy should vanish. Thus the Bekenstein-Hawking entropy must be the von Neumann entropy of another quantum state, presumably one that is obtained by an appropriate coarse-graining of the original state. What characterizes this coarse-grained state?

This question was the subject of a recent conjecture by Engelhardt and Wall (EW)~\cite{Engelhardt:2017aux}. The EW conjecture applies to a class of surfaces that may lie on or inside the event horizon. The Bekenstein-Hawking entropy associated with a ``minimar'' surface $\sigma$ is the area of the extremal (Ryu-Takayanagi~\cite{Ryu:2006bv} or HRT~\cite{Hubeny:2007xt}) surface, maximized over all spacetimes that agree with the given solution outside of $\sigma$. (The input spacetime may have no such surface and thus no entropy.) Engelhardt and Wall showed that the coarse-grained entropy so defined does indeed agree with the area of $\sigma$. The interpretation of extremal surface area as an entropy in the quantum gravity theory is well-motivated by the success of the RT proposal in asymptotically Anti-de Sitter spacetimes. We review the EW coarse-graining procedure in Sec.~\ref{sec-cew}. 

However, the EW construction and proof are purely classical. In particular, the construction fails when quantum matter is included, because it relies on the Null Energy Condition. Moreover, there is considerable evidence that in semi-classical gravity, it is the generalized entropy \cite{PhysRevD.7.2333} (and not the area) that 
is naturally associated with thermal states of the underlying quantum gravity theory~\cite{Faulkner:2013ana, Engelhardt:2014aa}.

Here, we will formulate a semi-classical extension of the EW coarse-graining proposal for black hole states; that is, we include effects that are suppressed by one power of $G\hbar$ compared to the classical construction. In Sec.~\ref{sec-calculation}, we consider a suitably defined quantum version of a ``minimar'' surface. At this order, we must hold fixed not only its exterior geometry but also the exterior state of the quantum fields. We conjecture a construction that explains the generalized entropy of the quantum minimar surface $\sigma$ in terms of a suitably coarse-grained state:
one can find an interior completion of the geometry and quantum state that contains a quantum stationary surface~\cite{Faulkner:2013ana, Engelhardt:2014aa, Dong:2017aa}
with equal generalized entropy, but none with larger generalized entropy. Moreover, we propose that saturation is obtained by extending $\sigma$ along a stationary null hypersurface whose classical and quantum expansions both vanish.

Unlike the classical EW construction, we cannot prove our conjecture. But in Sec.~\ref{sec-compare}, following the example of the QFC $\to$ QNEC derivation, we are able to extract a pure quantum field theory limit. We apply our construction to states on a fixed background black hole spacetime with a complete Killing horizon. In this limit, coarse-graining requires the existence of QFT states with specific and somewhat surprising properties, which we list. The most striking property of the coarse-grained state is that the energy flux across the horizon has delta-function support on $\sigma$; and that it vanishes at all earlier times on the horizon. (At later times the state agrees with the input state by construction.) The strength of the delta function is set by the derivative of the von Neumann entropy along the horizon in the input state, $\hbar S'/2\pi$.

In particular, the existence of a quantum state with these properties would imply a new result in QFT, Wall's ``ant conjecture''~\cite{Wall:2017aa} concerning the minimum energy of global completions of a half-space quantum state. (We review the ant conjecture in Appendix~\ref{sec:main}. The QNEC follows from this conjecture, but it has also been directly proven.) Our proposal thus implies that a state that maximizes the generalized entropy minimizes the nongravitational energy inside of a cut of a Killing horizon, subject to holding fixed the state on the outside. Roughly speaking, ignorance saves energy.

In fact, Wall's ant conjecture was recently proven by Ceyhan and Faulkner (CF)~\cite{Ceyhan:2018zfg}. The CF construction takes as input a state on a Killing horizon and a cut at some surface $\sigma$ on the horizon. Connes cocycle flow then generates a family of states that differ only to the past of the cut. In the limit of infinite flow, a state is approached whose properties prove the ant conjecture. 

In greater than 1+1 dimensions, the requirements we derive appear to be stronger than those demanded by the ant conjecture; see Appendix~\ref{sec:main}. Thus it is not immediately obvious that the quantum states required for our coarse-graining proposal exist. However, in Sec.~\ref{sec-cf} we show that the CF family of states attains all of the properties required by our conjecture. In particular, a delta function shock appears at the cut, with precisely the predicted strength. It is interesting that this feature arises in an algebraic construction whereas in the black hole setting, it arose geometrically from requiring a source for a discontinuity in the metric derivative. Thus, the CF construction proves the QFT limit of our conjecture, even though it was originally designed to prove the ant conjecture.



We briefly discuss some future directions in Sec.~\ref{sec-discussion}.

\section{Classical coarse-graining of black hole states}
\label{sec-cew}

In this section we review a classical geometric construction by Engelhardt and Wall (EW)~\cite{Engelhardt:2017aux,Engelhardt:2018kcs}. In Sec.~\ref{sec-cdefs}, we provide definitions of (classically) marginally trapped, ``minimar'', stationary, and HRT surfaces.

In Sec.~\ref{sec-EWclassical}, we summarize the EW proposal for the outer entropy of a ``minimar'' surface, a marginally trapped surface $\sigma$ that satisfies certain addition conditions. EW define this entropy in terms of geometries that agree with in the exterior of $\sigma$ but differ in the interior. For any such auxiliary geometry, inspired by the Ryu-Takayanagi proposal, the von Neumann entropy is assumed to be given by the area of a stationary surface. Maximizing this area over all possible auxiliary geometries, EW show that it agrees with the area of $\sigma$, which thus represents a coarse-grained entropy in agreement with the Bekenstein-Hawking formula.

\begin{figure}[]
\includegraphics[width=.32\textwidth]{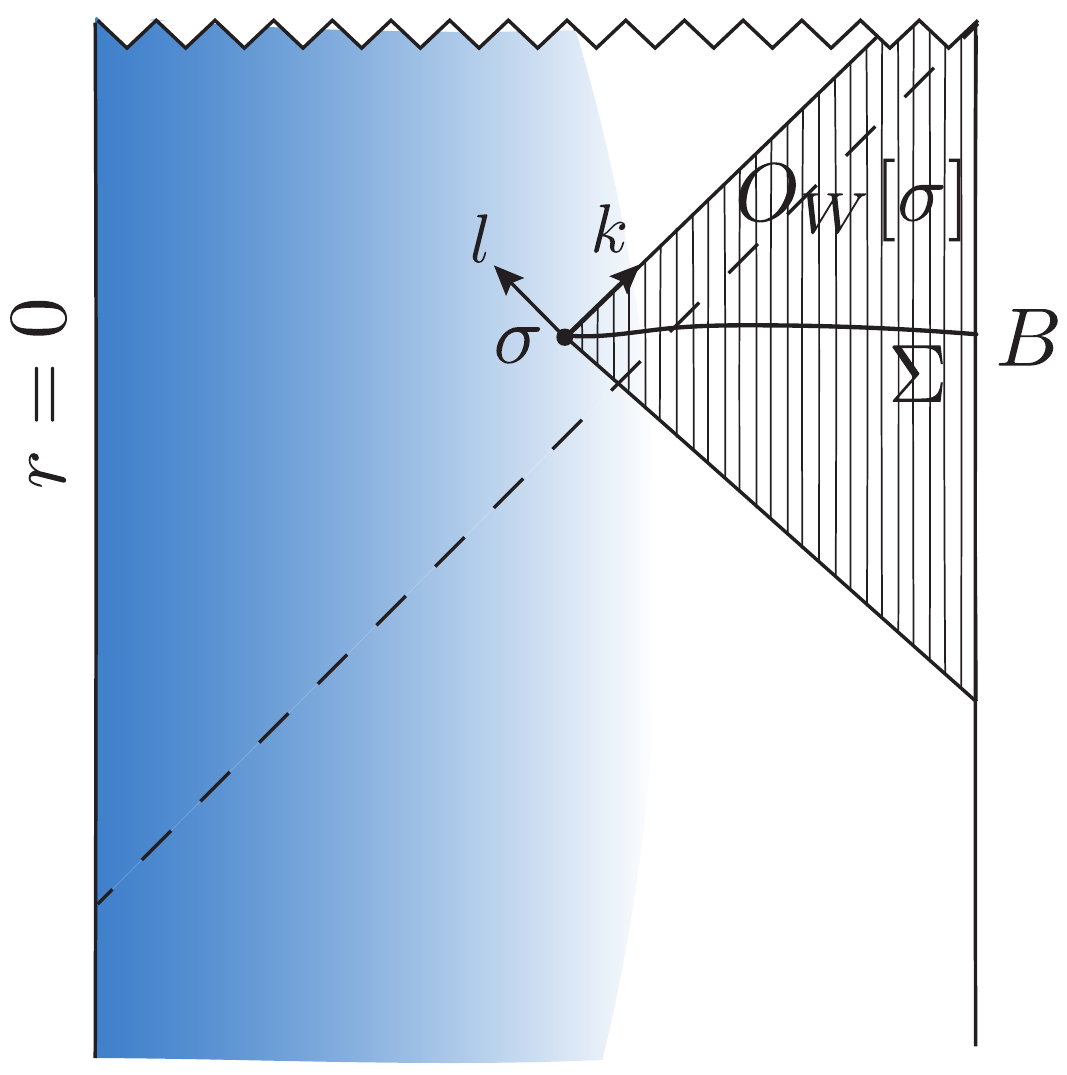}
\caption{Penrose diagram of a black hole formed from collapse in Anti-de Sitter space, showing a minimar surface $\sigma$ and its outer wedge $\mathcal{O}_{W}[\sigma]$ with Cauchy surface $\Sigma$.}
\label{fig1}
\end{figure}

\subsection{Classical marginal, minimar, and stationary surfaces}
\label{sec-cdefs}

We begin by fixing some notations and conventions; see Sec.~2 of~\cite{Engelhardt:2018kcs} for details. Let $\sigma$ be a {\em Cauchy splitting surface}, that is, $\sigma$ is an achronal codimension two compact surface that divides a Cauchy surface $\Sigma$ into two sides, $\Sigma_\text{in}$ and $\Sigma_\text{out}$. 

Let $k^a$, $l^a$ be the two future-directed null vector fields orthogonal to $\sigma$, normalized so that  $k_a l^a=-1$; and let $\theta_k$, $\theta_l$ be their expansions.

If exactly one null expansion vanishes, we shall take this to be the $k$-expansion. Then $\sigma$ is called {\em marginally outer trapped}, with $k$ defining the ``outside.'' If $\theta_l<0$ everywhere on a marginally outer trapped $\sigma$, we call $\sigma$ {\em marginally trapped}.

The {\em outer wedge} $O_W[\sigma]$ of a marginally trapped surface $\sigma$ is the set of spacelike separated events on the outside of $\sigma$ (the side that $k$ points towards, see above): $O_W[\sigma]\equiv D[\Sigma_\text{out}]$, where $D$ denotes the domain of dependence. See Fig.~\ref{fig1}. 

A {\em minimar surface} is a marginally trapped surface $\sigma$ that satisfies two additional restrictions:
\begin{itemize}
\item $O_W[\sigma]$ contains a connected component $B$ of an asymptotic conformal boundary (as would be the case, for example, if $\sigma$ lies in a single black hole formed from collapse in asymptotically anti-de Sitter or flat spacetime). Moreover, $O_W[\sigma]$ admits a Cauchy surface on which $\sigma$ is the surface homologous to $B$ that minimizes the area; see Fig.~\ref{fig1}.
\item $k^a\nabla_a \theta_{(l)} < 0$
\end{itemize}

A {\em stationary surface} $X$ is a surface whose expansion vanishes in both null directions, $k$ and $l$:
\begin{equation}
  \theta_k = \theta_l = 0 \text{ everywhere on } X~.
\end{equation}

A {\em Hubeny-Rangamani-Takayanagi (HRT) surface} $X$ is a stationary surface that satisfies additional requirements: it is the stationary surface with the smallest area, subject to a homology condition \cite{Hubeny:2007xt, Ryu:2006bv}. Here, we will require that $X$ be homologous to a minimar surface $\sigma$, and hence to a connected component $B$ of a conformal boundary.

\subsection{Bekenstein-Hawking entropy from coarse-graining behind minimar surfaces}
\label{sec-EWclassical}

Engelhardt and Wall~\cite{Engelhardt:2018kcs} argued that the area of a minimar surface $\sigma$ can be understood as a coarse-grained entropy. For geometries with a CFT dual, an explicit prescription for this coarse-graining can be formulated in the CFT. Here, we will be interested in the bulk definition of this coarse-graining, which can be discussed in more general geometries.

In the bulk, the coarse-graining consists of holding fixed the outer wedge of $\sigma$, $O_W[\sigma]$, while erasing the spatial interior of $\sigma$ and replacing it with an auxiliary geometry. One seeks the auxiliary geometry with the largest possible HRT surface $X$ behind $\sigma$. The coarse-grained entropy of $\sigma$ is defined as $A[X]/4G\hbar$.

So far, we have reviewed the definition of the outer entropy. The EW proposal is the conjecture that
\begin{itemize}
\item $S_\text{outer} \equiv A[X]/4G\hbar$ represents the von Neumann entropy of a well-defined state in a quantum gravity theory; and
\item $A[X]= A[\sigma]$.
\end{itemize}

EW proved the first part of the conjecture for the special case where $B$ lies on the conformal boundary of an asymptotically AdS spacetime, and $\sigma$ lies on a perturbed Killing horizon; moreover the proof assumes the Ryu-Takayanagi~\cite{Ryu:2006bv} and HRT~\cite{Hubeny:2007xt} proposals for the von Neumann entropy of the boundary CFT. In this case, it is possible to construct the dual CFT state explicitly, and to show that its entropy agrees with $S_\text{outer}$.

The second part of the conjecture was proven more generally~\cite{Engelhardt:2018kcs}. Using the maximin definition of the HRT surface~\cite{Wall:2012uf}, it can be shown that
\begin{equation}
  A[X]\leq A[\sigma]~.
  \label{eq-ineq}
\end{equation}
This argument assumes the Null Energy Condition (NEC), that the stress tensor satisfies
\begin{equation}
  T_{ab}k^ak^b \geq 0
  \label{eq-nec}
\end{equation}
for any null vector $k^a$.

EW explicitly construct an interior geometry that saturates the inequality (\ref{eq-ineq}). This implies
\begin{equation}
  S_\text{outer}[\sigma] \equiv \frac{A[X]}{4G\hbar} = \frac{A[\sigma]}{4G\hbar}~.
\end{equation}

The interior geometry with $A[X]=A[\sigma]$ is constructed by specifying initial conditions on the null hypersurface $N_k^-$ orthogonal to $\sigma$ towards the interior and past. Appropriate initial data is generated by null-translating the intrinsic geometry of $\sigma$, thus generating a stationary null hypersurface:
\begin{equation}
\theta_k = 0~~\text{on}~ N_k^- ~.
  \label{eq-tkis0}
\end{equation}
This ensures that all cross sections of $N_k^-$---in particular, $X$---have the same intrinsic metric and area as $\sigma$. This construction is consistent with the relevant constraint, the Raychaudhuri equation,
\begin{equation}
  k^a\nabla_a \theta_k = -\frac{1}{2} \theta_k^2 - \varsigma^2- 8\pi G\, T_{kk}~,
  \label{eq-raych}
\end{equation}
if one sets
\begin{equation}
  \varsigma=0 \text{ and } T_{kk}=0~~\text{on}~ N_k^- ~.
\label{eq-choices1}
\end{equation}
on $N_k^-$. EW~\cite{Engelhardt:2018kcs} show that this choice is always possible. Since $\theta_k$ vanishes on $\sigma$, Eqs.~(\ref{eq-raych}) and (\ref{eq-choices1}) ensure that the entire extrinsic curvature tensor in the $k$-direction vanishes everywhere on $N_k^-$, achieving the desired stationarity of $N_k^-$.

Moreover, it is important to show that there exists a stationary (HRT) surface $X$ on $N_k^-$. The outgoing expansion $\theta_k$ vanishes on any cut of $N_k^-$, by the above construction. The question is whether there exists a cut $X$ on which the ingong expansion $\theta_l$ vanishes as well. This is accomplished in the following sequence of steps.

The minimar assumption dictates that on $\sigma$, $\theta_l<0$ and $k^a\nabla_a\theta_l<0$. One can choose initial conditions on $N_k^-$ such that along every null generator of $N_k^-$, $k^a\nabla_a\theta_l$ is constant and equal to its value on $\sigma$: by the cross-focussing equation,
\begin{equation}
k^a\nabla_a \theta_l = -\frac{1}{2} {\cal R}  - \theta_k\theta_l +\chi^2+\nabla\cdot\chi + 8\pi G\, T_{kl}~,
\end{equation}
this can be accomplished by choosing all terms on the right hand side to be constant on $N_k^-$. This is already ensured for the intrinsic curvature scalar $\cal R$ and for the (vanishing) $\theta_k\theta_l$ term, by stationarity of $N_k^-$. The twist, or normal 1-form, is defined by
\begin{equation}
  \chi_a = h^c_{~a}l^d\nabla_c k_d~,
\end{equation}
where $h_{ab} = g_{ab}+2 l_{(a} k_{b)}$ is the induced metric on a cut. The twist evolves according to
\begin{equation}
  k^a\nabla_a \chi_i = 8\pi T_{ik} (+ \text{ terms that vanish when } \theta_k=\varsigma=0)~.
\end{equation}

To summarize, one can accomplish $k^a\nabla_a\theta_l =\left. k^a\nabla_a\theta_l\right|_\sigma$ on $N_k^-$ by choosing Eqs.~(\ref{eq-tkis0}) and (\ref{eq-choices1}) and in addition, along each null generator of $N_k^-$,
\begin{equation}
  T_{kl} = \left. T_{kl}\right|_\sigma \text{ and } T_{ik}=0~~\text{on}~ N_k^- ~.
  \label{eq-choices2}
\end{equation}
Again, EW argue that these choices are always possible.

Let $v$ be the affine parameter associated to $k^a$, and let $y$ be the transverse coordinates (angular coordinates) on $\sigma$. The location of a stationary surface $X$, $v=f(y)$, is determined by the differential equation
\begin{equation}
  L^a[f] = -\theta_l|_\sigma~,
  \label{eq-stability}
\end{equation}
where $L^a$ is the stability operator (see Ref.~\cite{Engelhardt:2018kcs} for details). This can be shown to have a solution with $-\infty<f<0$, so the HRT surface exists and lies on $N_k^-$. 

EW then glue the geometry exterior to $X$ (that is, $N_k^-$ and the outer wedge) to its CPT image across $X$. This constructs a ``two-sided'' geometry in which $X$ functions as a kind of bifurcation surface of a two-sided black hole/white hole pair. (However, the stationary auxiliary portion $N_k^-$ does not in general correspond to the horizon of a Kerr-Newman black hole, as its intrinsic metric can differ.)

In a final step, EW show that $X$ is not just stationary but is an HRT surface, i.e., that $X$ is the smallest-area stationary surface homologous to $\sigma$. This step uses the NEC as well as the second part of the minimar property of $\sigma$.

This concludes our summary of the EW coarse-graining prescription. Again, we refer the interested reader to Ref.~\cite{Engelhardt:2018kcs} for more detailed definitions and arguments.

\section{Semiclassical coarse-graining of black hole states}\label{sec-calculation}

In this section, we formulate a semiclassical extension of the Engelhardt-Wall construction, starting from a quantum marginally trapped surface $\sigma$. We conjecture that the semiclassical state invoked in our construction exists in the full quantum gravity theory; and that in this theory this state has a von Neumann entropy given by the generalized entropy of $\sigma$.

In Sec.~\ref{sec-qdefs}, we introduce relevant concepts such as generalized entropy, quantum expansion, quantum marginally trapped surfaces, and quantum HRT surfaces.

In Sec.~\ref{sec-EWquantum}, we state our quantum extension of the EW coarse-graining proposal.

In Sec.~\ref{sec-proof}, we refine our conjecture by describing key properties that the coarse-grained state is expected to satisfy at the level of semiclassical gravity. (These properties will be shown to have an interesting nongravitational limit in Sec.~\ref{sec-compare}. In Sec.~\ref{sec-cf} we will show that a recent construction by Ceyhan and Faulkner~\cite{Ceyhan:2018zfg} generates quantum field theory states which achieve these properties in a certain limit.)

\subsection{Quantum marginal, minimar, and stationary surfaces}
\label{sec-qdefs}

Before we turn to the question of why and how the EW construction should be extended to the semiclassical regime, we introduce here the relevant concepts: generalized entropy, quantum expansion, quantum (marginally) trapped surfaces, and quantum extremal surfaces. More details can be found, e.g., in Refs.~\cite{C:2013uza,
  Engelhardt:2014gca,Bousso:2016aa,Wall:2011hj}. 

The notion of {\em generalized entropy} was originally introduced by Bekenstein~\cite{PhysRevD.7.2333} as an extension of ordinary entropy that includes the contribution from black holes, $S_{\rm out} \to S_{\rm out} +\frac{A}{4G\hbar}$. But in an expansion in $G\hbar$, it is the exterior entropy that should be regarded as a quantum correction:
\begin{equation}
  S_{\rm gen} = \frac{A}{4G\hbar} + S_{\rm out}+ \ldots~,
\end{equation}
Equivalently, $4G\hbar S_{\rm gen}$ represents a quantum-corrected area.

In Bekenstein's original proposal, $A$ represented the area of a cut of a black hole event horizon; and $S_{\rm out}$ represented the entropy in the black hole's exterior. However, the generalized entropy can be defined for any Cauchy-splitting surface $\sigma$, with $S_{\rm out}$ the von Neumann entropy of the quantum fields restricted to one side of $\sigma$. $A/4G\hbar$ should be regarded as the leading counterterm that cancels divergences in the entropy; we suppress subleading terms here. Given its wide applicability, the notion of generalized entropy can be used to define quantum-corrected notions of trapped, stationary, etc., as follows.

Recall that the classical expansion of a surface $\hat\sigma$ at a point $y\in\hat\sigma$ is the trace of the null extrinsic curvature at $y$. It can also be defined as a functional derivative,
\begin{equation}
\theta[\hat\sigma;y] = h(y)^{-1/2} \frac{\delta A[V]}{\delta V(y)}~,
\end{equation}
where $h$ is the area element on $\hat\sigma$. Here $V(y)$ defines a surface that lies an affine parameter distance $V$ from $\hat\sigma$ along the null geodesic emanating from $\hat \sigma$ at $y$.

The above definition is overkill, as the classical expansion depends only on the local geometry near $y$. But it generalizes directly to the {\em quantum expansion}, $\Theta$, which depends on $\hat\sigma$ nonlocally:
\begin{equation}
  \Theta[\hat\sigma;y] = \frac{4G\hbar}{\sqrt{h(y)}} \frac{\delta S_{\rm gen}[V]}{\delta V(y)}~.
\end{equation}

A {\em quantum marginally outer trapped surface} is a surface whose quantum expansion in one of the two null directions (say, $k$) vanishes at every point. Let $\sigma$ be such a surface:
\begin{equation}
  \Theta_k[\sigma;y] \equiv 0~.
\end{equation}
It follows that
\begin{equation}
  \theta_k(y) = -\frac{4G\hbar}{\sqrt{h(y)}} \frac{\delta S_{\rm out} }{\delta V(y)}
  \label{eq-bal}
\end{equation}
at every point on $\sigma$.

A {\em quantum marginally trapped surface} is a quantum marginally outer trapped surface for which in addition
\begin{equation}
  \Theta_l[\sigma;y] < 0 ~.
\end{equation}
(As usual, {\em anti-trapped} corresponds to the opposite inequality on the $l$-expansion.)

The {\em outer wedge} $O_W[\sigma]$ of a quantum marginal surface $\sigma$ is the set of spacelike separated events on the ``marginal'' side of $\sigma$, i.e., the side that $k$ points towards: $O_W[\sigma]=D[\Sigma_\text{out}]$; see Fig.~\ref{fig1}. 

A {\em quantum minimar surface}, is a quantum marginally trapped surface $\sigma$ that satisfies two additional restrictions:
\begin{itemize}
\item $O_W[\sigma]$ contains a connected component of an asymptotic conformal boundary (as would be the case, for example, if $\sigma$ lies in a single black hole formed from collapse in asymptotically anti-de Sitter or flat spacetime). Moreover, $O_W[\sigma]$ admits a Cauchy surface on which $\sigma$ is the surface homologous to $B$ that minimizes the generalized entropy; see Fig.~\ref{fig1}.
\item $k^a\nabla_a \theta_l < 0$~.
\end{itemize}
Note that we impose the second condition on the classical expansion, not the quantum expansion. Since the inequality is strict, the classical expansion $\theta_l$ will dominate in the semiclassical expansion in $G\hbar$.

A {\em quantum stationary surface}\footnote{In an abuse of language, this is sometimes referred to as extremal rather than stationary.} is a surface whose quantum expansions vanish in both null directions, $k$ and $l$. We will demand that $X$ be such a surface:
\begin{equation}
  \Theta_k[X;y] \equiv 0~,~~\Theta_l[X;y] \equiv 0~.
\end{equation}

A {\em quantum HRT surface} satisfies additional requirements: it is the quantum stationary surface with the smallest generalized entropy; and it must obey a homology condition. Here, we will require that it be homologous to a quantum minimar surface $\sigma$, and and hence to a connected component $B$ of a conformal boundary.

\subsection{Generalized entropy from coarse-graining behind quantum marginally trapped surfaces}
\label{sec-EWquantum}

We will now motivate and formulate a quantum extension of the EW proposal. To see that such an extension is needed, note that the classical EW construction relies on the Null Energy Condition, Eq.~(\ref{eq-nec}). The NEC guarantees that no HRT surface with area greater than that of the marginally trapped surface can be constructed. It also guarantees that the stationary surface with equal area is an HRT surface. But the NEC is known to fail in any relativistic quantum field theory, so none of these conclusions survive at the semiclassical level.

Indeed, one does not expect any quantum state of the full quantum gravity theory to correspond to just the area of a surface (as is implicit in the classical EW construction). Rather, one expects its von Neumann entropy to match the generalized entropy. That is, to the extent that a quantum state corresponds to a surface, one expects it to also describe the surface's exterior.

There is significant evidence supporting this expectation from the AdS/CFT correspondence~\cite{Maldacena:1997re}. Consider the quantum state $\rho_B$ on a region $B$, where $B$ can be all or part of the boundary. This state is expected~\cite{Faulkner:2013ana} to describe the entire {\em entanglement wedge} of $B$, i.e., the spacetime region enclosed by $B$ and the HRT surface $X[B]$. The $1/N$ expansion on the boundary (with $N$ the rank of the CFT's gauge group) corresponds to the $G\hbar$ expansion in the bulk. In particular, the von Neumann entropy $S(\rho_B)$ can be expanded in this way, with the leading $O(N^2)$ piece corresponding to the area of $X[B]$, and the subleading $O(1)$ piece corresponding to the exterior bulk entropy $S_{\rm out}$. When expanding to higher orders, $X_B$ should be taken to be the quantum HRT surface of $B$~\cite{ Engelhardt:2014gca}.

We thus seek a proposal in which the generalized entropy of a surface $\sigma$ is explained as a coarse-grained entropy. The coarse-graining should correspond to maximizing the generalized entropy of a quantum HRT surface $X$, subject to holding fixed the outer wedge $O_W[\sigma]$ (now including the quantum state of bulk fields in $O_W[\sigma]$). The coarse-graining prescription will be successful if $S_{\rm gen}[X]=S_{\rm gen}[\sigma]$.

The remaining question is what characterizes a surface $\sigma$ that we may consider for coarse-graining. In the classical case, the appropriate criterion was that $\sigma$ be minimar. In the quantum case, the natural candidates are minimar surfaces or quantum minimar surfaces. In the EW construction of the maximally coarse-grained state, the HRT surface $X$ of the coarse-grained state lies on a stationary null surface $N_k^-$ extended to the past and inwards from $\sigma$. Our construction will share this feature. This excludes (classically) minimar as the relevant criterion for $\sigma$. The variation of $S_{\rm out}$ does not have definite sign on such surfaces, and so their quantum expansion would not have a definite sign. However, if $\Theta[\sigma]>0$ then by the quantum focussing conjecture, it would be impossible to find an $X$ with $\Theta[X]=0$ on $N_k^-[\sigma]$. Therefore, we will require that $\sigma$ be quantum minimar; in particular, $\Theta[\sigma]=0$. 

We now state our {\bf proposal}. Let $\sigma$ be a quantum minimar surface homologous to a boundary region $B$, with generalized entropy $S_{\rm gen}[\sigma]$ and outer wedge $O_W[\sigma]$. Let $\bar X$ be a quantum HRT surface in any geometry such that:
\begin{itemize}
\item $O_W[\bar X]\supset O_W[\sigma]$.
\item $\bar X$ is homologous to $\sigma$.
\item Both the geometry and the quantum state of $O_W[\bar X]$ agree with that of $O_W[\sigma]$ upon restriction of $O_W[\bar X]$ to $O_W[\sigma]$. (To be precise, let $\Sigma_\text{out}[\bar X]$ be a Cauchy surface of $O_W[\bar X]$ such that $\Sigma_\text{out}[\bar X]\cap O_W[\sigma]$ is a Cauchy surface of $O_W[\sigma]$, $\Sigma_\text{out}[\sigma]$, and let $\rho_{\bar X}$ and $\rho_\sigma$ be the state of the quantum fields on $\Sigma_\text{out}[\bar X]$ and $\Sigma_\text{out}[\sigma]$, respectively. We require that $\text{Tr}_{\Sigma_\text{out}[\bar X]-\Sigma_\text{out}[\sigma]}\, \rho_{\bar X}=\rho_\sigma$.)
\end{itemize}
We claim that
\begin{equation}
  \text{sup}_{\bar X} S_\text{gen} [\bar X] = S_\text{gen}[\sigma]~.
  \label{eq-conjecture}
\end{equation}
Moreover, let $X$ be a surface $\bar X$ that achieves the supremum. (This should be taken as a limiting statement if no such $X$ exists.) Then $O_W[X]$ represents a coarse-graining of the original geometry, with respect to the quantum minimar surface $\sigma$. In particular, in AdS/CFT the quantum state on $B$ dual to the entanglement wedge $O_W[X]$ has von Neumann entropy $S_\text{gen}[\sigma]$.

Unlike the classical case, we will not prove this conjecture, but we will provide some evidence supporting its plausibility. We proceed in two steps as in the classical case: first, we will argue that
\begin{equation}\label{upperboundSX}
  S_\text{gen} [\bar X] \leq S_\text{gen}[\sigma]
\end{equation}
for any $\bar X$ satisfying the conditions in our proposal. We then refine our conjecture by detailing the properties of a semiclassical geometry and quantum state that would achieve equality.

In order to show Eq. \eqref{upperboundSX}, we generalize the result in \cite{Engelhardt:2018kcs} to the quantum case. This involves two main assumptions. The first assumption is the quantum focusing conjecture~\cite{Bousso:2015mna} which asserts that in the semi-classical limit the derivative of the quantum expansion of codimension 2 surfaces under any null deformation is non-negative:
\begin{align}
  \frac{\delta \Theta_{k}[X; y]}{\delta V(y)}  \leq 0~.
\end{align}
The second assumption is a slightly weaker quantum generalization of the classical maximin construction \cite{Wall:2012uf}. More precisely, we assume that the quantum extremal surface $\bar X$ is also the surface of minimal generalized entropy on some Cauchy slice $\Sigma$.

By global hyperbolicity, the congruence of null geodesics orthogonal to $\sigma$ in the $\pm k$ directions intersect $\Sigma$ at some Cauchy splitting surface $\bar{\sigma}$. (The congruence should be terminated at conjugate points or self-intersections~\cite{Bousso:2015mqa,Akers:2017nrr}. Since $\sigma$ is a quantum marginally trapped surface, quantum focusing ensures that
\begin{align}
  S_\text{gen} [\bar \sigma] \leq S_\text{gen}[\sigma]~.
\end{align}
The quantum maximin assumption further implies
\begin{align}
  S_\text{gen} [\bar X] \leq S_\text{gen}[\bar \sigma]~,
  \label{eq-bound}
\end{align} 
which establishes Eq. \eqref{upperboundSX}.

\subsection{Properties of a Generalized Entropy Maximizing Bulk State}
\label{sec-proof}


We will now describe a geometry and quantum state with a quantum extremal surface $X$ whose generalized entropy saturates the inequality (\ref{upperboundSX}). The existence of a state with the properties we describe would imply our conjecture, Eq.~(\ref{eq-conjecture}).

By asserting the existence of this semiclassical state, we are refining our conjecture. In Sec.~\ref{sec-compare}, we will explore the implications of this refinement in a pure field theory limit. In Sec.~\ref{sec-cf}, we will show that these implications are realized in a recent construction by Ceyhan and Faulkner~\cite{Ceyhan:2018zfg}.

Our construction will be analogous to the classical one, in that we will approach $X$ along the null hypersurface $N_k^-[\sigma]$. Since we require $\Theta_k[\sigma]=\Theta_k[X]=0$, the quantum focussing conjecture ($\Theta_k'\leq 0$) requires that $\Theta_k=0$ everywhere on $N_k^-$.  That is, $S_{\rm gen}$ must be constant along $N_k^-$. (This is analogous to classical focussing and the null energy condition requiring that $N_k^-$ have constant area in the classical case.) 

In the classical case, all relevant quantities could be chosen to be constant on $N_k^-$. In other words, the surface $N_k^-$ is truly stationary. This would not be the case if $\theta$ and the derivative of the entropy varied along $N_k^-$, with only their sum $\Theta_k$ vanishing. Motivated by this observation, we conjecture that a state can be found such that the two terms in $\Theta_k$ vanish {\em separately} on $N_k^-$:
\begin{equation}
  \theta_k=0 ~~~\text{and}~~~\frac{\delta S_{\rm out} }{\delta V(y)}=0~.
  \label{eq-strong}
\end{equation}
In analogy with the classical construction we also take the shear tensor to vanish at all orders in $\hbar$ along $N_k^-$:
\begin{equation}
  \varsigma= 0 ~.
\end{equation}

These considerations place nontrivial constraints on the limit state we seek. For $\theta_k$ and $\varsigma$ to vanish everywhere on $N_k^-$, the stress tensor component $T_{kk}$ must vanish on $N_k^-$. Moreover, note that $\theta_k$ need not vanish on $\sigma$, where only $\Theta_k=0$ is required. It follows that generically, $\theta_k$ must jump discontinuously, by an amount
\begin{equation}
  \Delta \theta_k|_\sigma = -\frac{4G\hbar}{\sqrt{h(y)}}
  \left . \frac{\delta S_{\rm out} }{\delta V(y)} \right|_\sigma~.
\end{equation}
By Raychaudhuri's equation, this implies the presence of a delta function term in the stress tensor, at $\sigma$. Combining these results, we conclude that
\begin{equation}
  T_{vv} = \frac{\hbar}{2\pi} \left. \frac{\delta S_{\rm out} }{\delta V(y)} \right|_\sigma \delta(v)~~,v\leq 0~, \label{shock}
\end{equation}
i.e., in the region $N_k^- \cup \sigma$.

To summarize, we conjecture the existence of a state with
\begin{eqnarray} 
  T_{vv} &  = &  \frac{\hbar}{2\pi} \left. \frac{\delta S_{\rm out} }{\delta V(y)} \right|_\sigma \delta(v)~~,~v\leq 0~, \label{eq-Tvv0} \\
  \varsigma & = & 0~~,~v<0~,\\
  \frac{\delta S_{\rm out} }{\delta V(y)} & = & 0~~,~v < 0.
\end{eqnarray}
Eq.~(\ref{eq-Tvv0}) trivially implies that
\begin{equation}
  \int_{-\infty}^v dv\, T_{vv}=0~,
  \label{eq-intTvv0}
\end{equation} 
and we will use this property in Sec.~\ref{sec-compare}.\footnote{Strictly, we must allow for the possibility that a state with the properties we conjecture does not itself exist. It suffices that the properties we require can be arbitrarily well approximated by some family or sequence of states (as in the example of Sec.~\ref{sec-cf}). In this case, Eq.~(\ref{eq-Tvv0}) need not imply Eq.~(\ref{eq-intTvv0}), so the latter property should be considered explicitly as part of our refined conjecture.} In addition, we assume that the remaining EW conditions listed in Eq.~(\ref{eq-choices2}) can be met at the classical level.

With these assumptions, the existence of a classical HRT surface on $N_k^-$ is guaranteed by the argument summarized around Eq.~(\ref{eq-stability}). This surface satisfies $\theta_l=0$. A quantum stationary surface $X$ can be found nearby (in the $G\hbar\to 0$ limit), by solving iteratively for $\theta_l=-\frac{4G\hbar}{\sqrt{h(y)}} \frac{\delta S}{\delta U(y)}$, where the functional derivative refers to the shape deformation along the $l$-congruence.

Finally, we need to show that $X$ is quantum HRT, i.e., that it is the quantum stationary surface homologous to $\sigma$ with smallest generalized entropy. This proceeds in exact analogy with the classical argument~\cite{Engelhardt:2018kcs}, with the QFC replacing the NEC, so we will not spell out the argument here. See \cite{Engelhardt:2019hmr} for details.

\section{Quantum field theory limit of coarse-grained quantum gravity states}
\label{sec-compare}

\begin{figure}[]
\includegraphics[width=.30\textwidth]{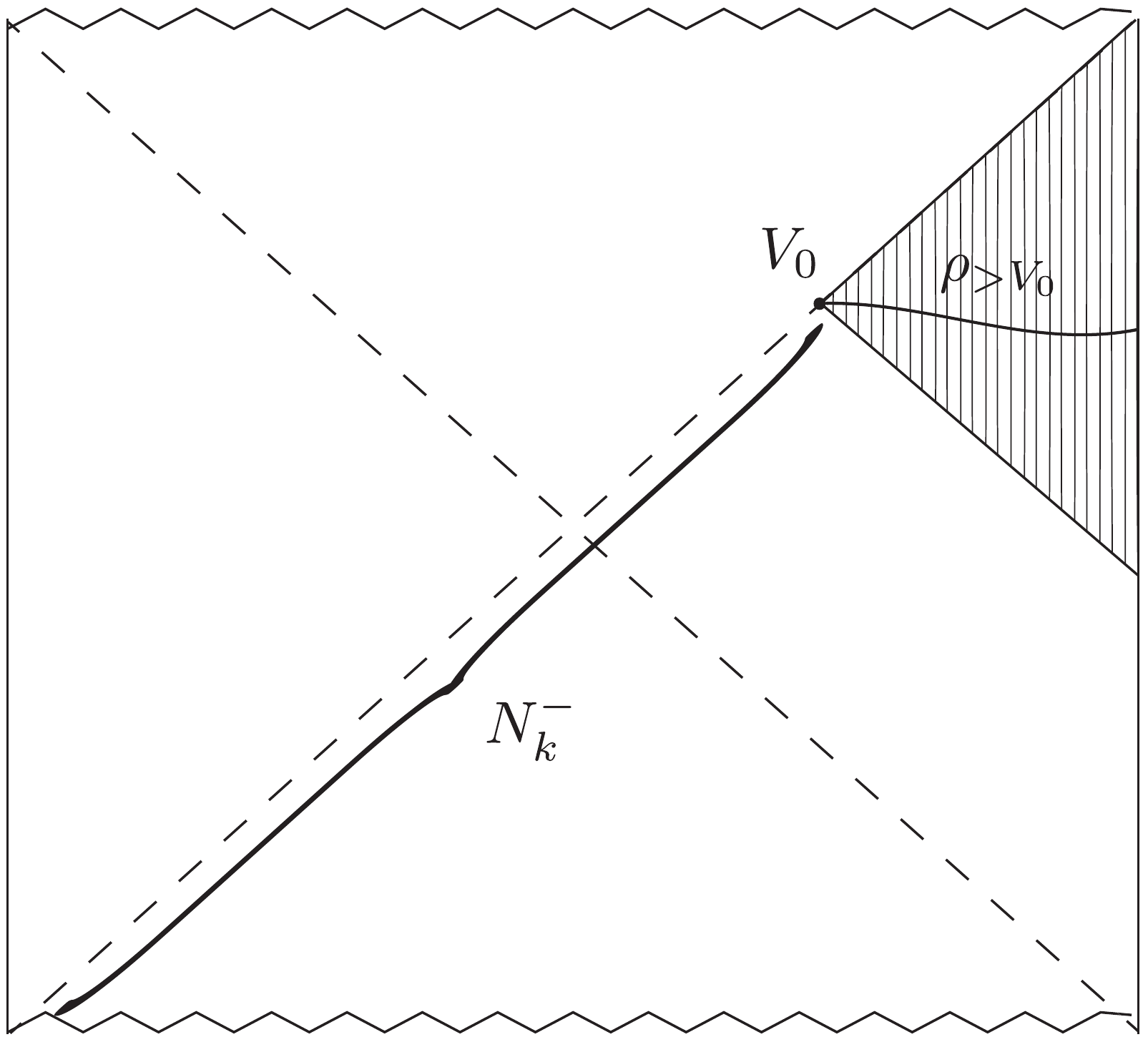}
\caption{Coarse-graining behind a Killing horizon. Any cut $V_0$ can be viewed as a quantum marginally trapped surface in the limit as $G\to 0$. The state $\rho_{>V_0}$ on the Cauchy surface $\Sigma$ of the outer wedge is held fixed. The coarse-grained geometry is the original geometry. The stationary null surface $N_k^-$  is the past of $V_0$ on the Killing horizon. The coarse-grained quantum state demanded by our proposal lives on $N_k^- \cup \sigma\cup\Sigma$. We identify the properties the state must have, and we show that the Ceyhan and Faulkner ``ant states'' satisfy these.}
\label{fig2dant}
\end{figure}

In this section, we study the implications of our conjecture for quantum field theory decoupled from gravity. We will apply our proposal to input states that are small perturbations of the Killing horizon of a maximally extended vacuum solution such as Kruskal; see Fig.~\ref{fig2dant}.

In the perturbative setting, any quantum marginally trapped surface $\sigma$ will be at a distance of order $G$ from the Killing horizon, and so will lie on the horizon as $G\to 0$. We can think of the area and null expansion of $\sigma$ as fields defined on the unperturbed Killing horizon whose changes are sourced by the state of the matter fields on the horizon. Thus, every cut of the Killing horizon can be viewed as quantum marginally trapped, and our conjecture can be applied.


We will first establish notation and review some standard results in Sec.~\ref{sec-not}. In Sec.~\ref{sec-compare}, we will derive some interesting additional properties of the coarse-graining states that must hold in the perturbative setting. In the limit as $G\to 0$, our conjecture thus implies the existence of states with both the properties established in the previous section, and the additional properties derived here, in quantum field theory on a fixed background. This is an in-principle testable conjecture about quantum field theory.

\subsection{Notation, definitions, and standard results}
\label{sec-not}

Consider a quantum field theory on a background with a Killing horizon 
and an arbitrary global state $\rho$ defined on the horizon. Let $v$ be the affine parameter on the Killing horizon, $u$ the affine parameter that moves off of the Killing horizon (associated with null vectors $k$ and $l$ respectively), and take $y$ to be the transverse coordinates on a cut $V(y)$ of the horizon. The cut defines a surface $\sigma$, which we assume to be Cauchy-splitting as usual.

Let the {\em right half-space state} $\rho_{>V_0}$ be the restriction of $\rho$ to the half-space $v>V_0(y)$ as in Fig. \ref{fig2dant}:
\begin{equation}
  \rho_{>V_0}\equiv \text{Tr}_{\leq V_0} \rho~.
\end{equation}
where the trace is over the algebra associated with the complement region.
Let us denote the {\em von Neumann entropy} of $\rho_{>V_0}$ by
\begin{equation}
  S(V_0) = - \text{Tr}\, \rho_{>V_0}\log\rho_{>V_0}~.
\end{equation}
Let $\sigma\equiv |\Omega\rangle\langle\Omega|$ be the global vacuum, which can be reduced to the {\em right vacuum} $\sigma_{>V_0}=\text{Tr}_{\leq V_0} \sigma$. The {\em vacuum-subtracted von Neumann entropy} of $\rho_{>V_0}$ is
\begin{equation}
  \Delta S(V_0) = S(V_0) + \text{Tr}\, \sigma_{>V_0}\log\sigma_{>V_0}~.
\end{equation}
The {\em right (half-)modular Hamiltonian} $K$ is defined by the relation
\begin{equation} \label{Modham}
\sigma_{>V_0} = \frac{e^{-K(V_0)}}{\text{Tr}\,e^{-K(V_0)}}~.
\end{equation}
The {\em right modular energy} in a global state $\rho$ is $\langle K(V_0) \rangle\equiv \text{Tr}\, [K(V_0) \rho_{>V_0}]$, and the {\em vacuum-subtracted right modular energy} is
\begin{eqnarray} 
  \Delta K (V_0) &\equiv &
                           \langle K(V_0)\rangle - \text{Tr}\, [\sigma_{>V_0} K(V_0)]  \\
                 & = & \frac{2\pi}{\hbar}\int dy \int_{V_0(y)}^\infty dv\, [v-V_0(y)] T_{vv}~,
                       \label{eq-k11}
\end{eqnarray} 
where the explicit expression is due to Bisognano and Wichmann~\cite{Bisognano:1975ih} and its generalization to arbitrary cuts of Killing horizons \cite{Faulkner:2016mzt, Casini:2017roe}.
The {\em relative entropy} of $\rho_{>v_0}$ with respect to the reduced global vacuum, $\sigma_{>v_0}$, is defined as
\begin{eqnarray}
  S_\text{rel}(V_0) & \equiv  & S(\rho_{>V_0}|\sigma_{>V_0})\\ & \equiv & \text{Tr}\, \rho_{>V_0}\log\rho_{>V_0}- \text{Tr}\, \rho_{>V_0}\log\sigma_{>V_0}~.
\end{eqnarray}
It follows from this definition that
\begin{equation}
  S_\text{rel}(V) = \Delta K(V) - \Delta S(V)~.
  \label{eq-srelks}
\end{equation}
We will often be interested in derivatives, where the vacuum-subtraction drops out. For example, 
\begin{equation}
  \frac{\delta K}{\delta V(y)}  =\frac{\delta  \Delta K}{\delta V(y)} = -\frac{2\pi}{\hbar} \int _v^\infty d\tilde v~ T_{vv}(y) ~.
  \label{eq-pvk}
\end{equation}
Similar definitions apply to the region $v<V_0$; we denote the associated ``left'' quantities with an overbar. Strictly, we define the left and right quantities in terms of the limit as $\epsilon\to 0$ of the open intervals $(-\infty, V_0(y)+\epsilon)$ and $(V_0(y)+\epsilon,\infty)$, respectively. The small shift ensures that any distributional sources at $V_0(y)$ contribute asymmetrically to the left but not to the right quantities. (We will see that in the minimum energy states of interest in this paper, the stress tensor generically has a delta function at $V_0(y)$. Our choice resolves an associated ambiguity, attributing this energy entirely to the left.)

\begin{figure}[]
\includegraphics[width=.38\textwidth]{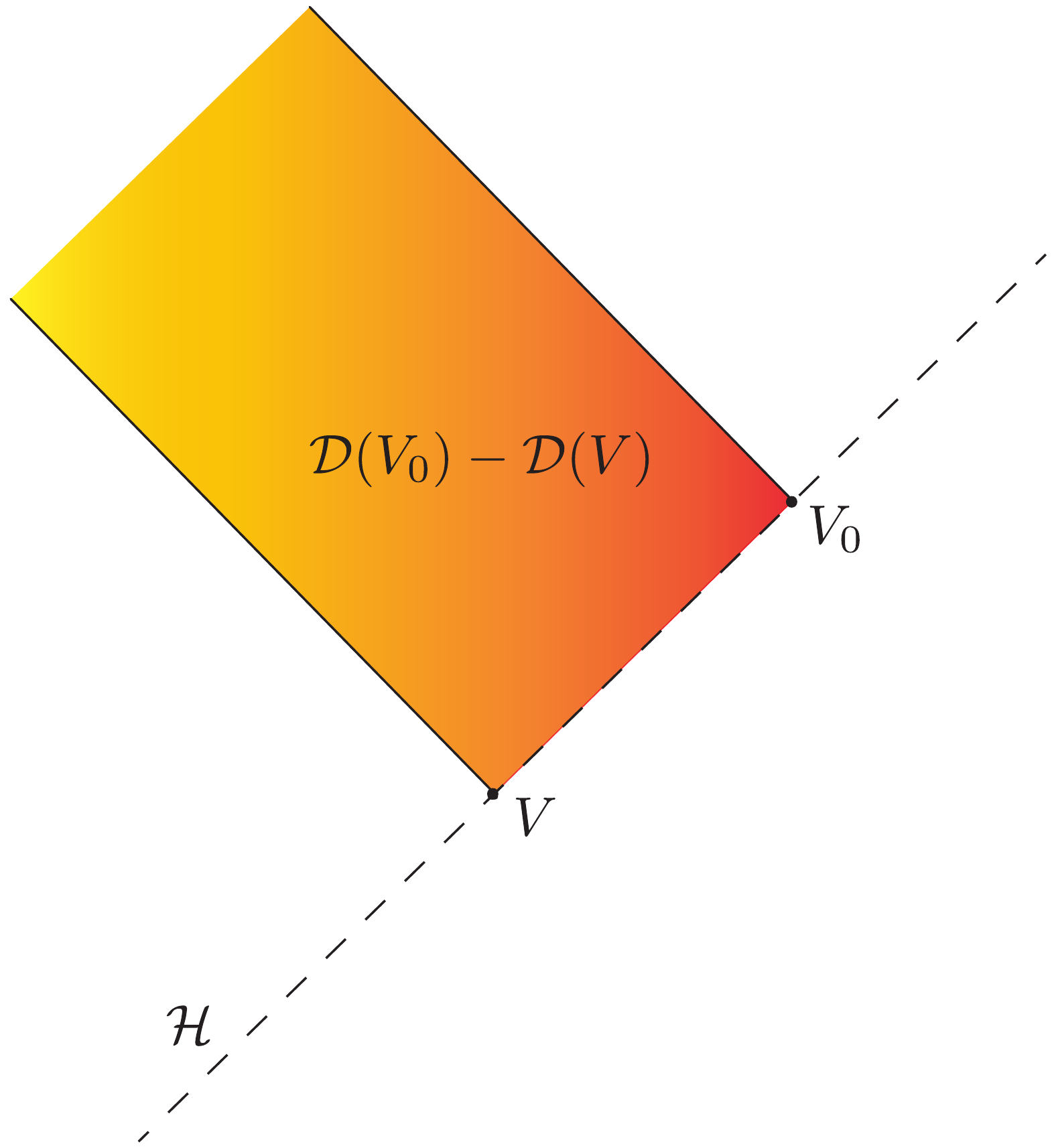}
\caption{The spacetime region associated to the interval $V < v < V_0$ on the null surface for which all observables in the algebra should register vacuum values in the coarse-graining state. \label{figdomains}}
\end{figure}

The relative entropy satisfies positivity and monotonicity:
\begin{equation}
S_\text{rel} \geq 0~,~~ \frac{\delta S_\text{rel}}{\delta V} \leq 0~.
\end{equation}
Via Eq.~(\ref{eq-srelks}), monotonicity implies
\begin{align}
  \frac{\delta \bar K}{\delta{V}} \geq \frac{\delta \bar S}{\delta V} \geq \frac{\delta S}{\delta V}~.
  \label{eq-kss}
\end{align}
The second inequality follows from the strong subadditivity of the von Neumann entropy,
\begin{equation}
  S_{BC}+S_{CD}\geq S_B+S_D~,
\end{equation}
applied to the intervals $B= (-\infty,v_0), C=[v_0,v_0+\delta], D=(v_0+\delta,\infty)$ in the limit as $\delta\to 0$~\cite{Wall:2017aa}.

\subsection{Additional properties of the coarse-graining states}

Our conjecture says that the coarse-grained state will have vanishing $T_{vv}$ and constant right entropy in the left region:
\begin{eqnarray} 
  \langle T_{vv} \rangle &  = &  \frac{\hbar}{2\pi} \left. \frac{\delta S }{\delta V(y)} \right|_\sigma \delta(v-V_{0}(y))~~,~v\leq V_{0}~,
  \label{eq-tvvkh}\\
  \frac{\delta S }{\delta V(y)} & = & 0~~,~v < 0~.
                                      \label{eq-sprimekh}
\end{eqnarray}
In particular, in the strong form of Eq.~(\ref{eq-intTvv0}), these properties imply the ant conjecture (see Appendix~\ref{sec:main}).

But additionally, on the Killing horizon, the nested inequalities \eqref{eq-kss} hold. Combined with the above equations, this implies that the left von Neumann entropy is also constant:
\begin{align}
&0=\int_{-\infty}^{V(y)} \langle T_{vv}\rangle \geq \frac{\delta \bar{S}}{\delta V(y)} \geq \frac{\delta S}{\delta V(y)}=0 \\
  &\implies \frac{\delta \bar{S}}{\delta V(y)}  =0, ~ v < V_{0}(y)~.
    \label{eq-sbaradd}
\end{align}
By Eqs.~(\ref{eq-intTvv0}), (\ref{eq-srelks}) and (\ref{eq-pvk}), it follows that the left relative entropy is constant:
\begin{align}\label{Srelprimezero}
  \frac{\delta \bar{S}_{\text{rel}}(\rho_{<V}|\sigma_{<V})}{\delta V(y)} = 0, ~ v < V_{0}(y)~.
\end{align}

But the relative entropy is a measure of the distinguishability of the state $\rho_{<V}$ from the vacuum $\sigma_{<V}$. Suppose that by moving up the cut $V$, i.e., by gaining access to a larger region, one could perform some measurement that would better distinguish $\rho_{<V}$ from the vacuum. Then the relative entropy of the larger region would have to be greater. Thus, Eq.~(\ref{Srelprimezero}) implies that all observables restricted to the difference between the left domains of dependence associated to cuts $V_{0}(y)$ and $V(y)$ (as in Fig. \ref{figdomains}) need to register vacuum values. In particular, the stress tensor one-point function must vanish:
\begin{align}\label{Tmunuzero}
\langle T_{\mu\nu}(x) \rangle = 0, ~~ x \in \mathcal{D}(V_{0})- \mathcal{D}(V)~.
\end{align}

It is more subtle to draw conclusions about $\langle T_{\mu\nu}(x) \rangle$ when $x$ is on the boundary of the region (marked by red in Fig. \ref{figdomains}), 
$u=0, v<V_{0}$. Because $T_{\mu\nu}$ does not exist as an operator unless it is smeared to both sides of this boundary, it will not be in the left operator algebra, and it cannot be used to distinguish $\rho_{<V}$ from the vacuum $\sigma_{<V}$. 

We will now give a rough physical argument that certain components of $\langle T_{\mu\nu}(x) \rangle$ must vanish also on the Killing horizon below the cut, $u=0, v<V_{0}$. We emphasize that this argument is not rigorous, as it borrows from classical intuition. (In forthcoming work we will explore a more detailed coarse-graining proposal involving a family of states; in that setting a rigorous argument can be given.)

Physically, $\langle T_{vv} \rangle$ can be thought of as the momentum orthogonal to an observer's worldline in the $(u,v)$ plane, in the limit as the observer moves at the speed of light in the $v$-direction. Similarly, $T_{iv}$ is the transverse momentum seen by such an observer. Since all observables in the algebra associated to $\mathcal{D}(V_0)- \mathcal{D}(V)$ have to register vacuum values, no excitations can enter this region. By causality, therefore, the state on the null surface $u=0, v<V_{0}$ can only differ from the vacuum by matter moving \emph{along} it, i.e., purely in the $v$-direction. This implies $\langle T_{vv} \rangle=0$, consistent with Eq.~(\ref{eq-sprimekh}) above. It also implies the new result
\begin{equation}
  \langle T_{iv}\rangle = 0~,~~v<V_0~.
  \label{eq-tivadd}
\end{equation}
Conservation of the stress tensor, 
\begin{align}
-\partial_{v} \langle T_{uv} \rangle -\partial_{u} \langle T_{vv} \rangle+\partial_{i} \langle T_{i v} \rangle = 0~,
\end{align}
combined with \eqref{eq-tvvkh} then yields
\begin{equation}
  \langle T_{uv}\rangle = \text{const}~.
  \label{eq-tuvadd}
\end{equation}

We conclude that coarse-grained states on Killing horizons must satisfy not only Eqs.~(\ref{eq-tvvkh}) and (\ref{eq-sprimekh}) but also Eqs.~(\ref{eq-sbaradd}), (\ref{Srelprimezero}), (\ref{eq-tivadd}), and (\ref{eq-tuvadd}).

Crucially, these results pertain to quantum field theory on a fixed background, so they can be checked in a rigorous setting. In the next section we will see that all of the above properties are indeed satisfied by the ``ant states'' constructed by Ceyhan and Faulkner~\cite{Ceyhan:2018zfg}. This proves our conjecture in the Killing horizon limit. 


\section{Existence of coarse-graining states in QFT limit}
\label{sec-cf}
  
In this section we show that the ``predictions'' of the previous section have already been confirmed. We consider a recent explicit construction of states in QFT by Ceyhan and Faulkner (CF)~\cite{Ceyhan:2018zfg}. CF constructed these states in order to prove a conjecture by Wall \cite{Wall:2017blw} that we will discuss in detail in Appendix~\ref{sec:main} below. For now, we merely verify that they satisfy the properties we found for the coarse-graining state on Killing horizons in the non-gravitational limit: Eqs.~(\ref{eq-tvvkh}), (\ref{eq-sprimekh}), (\ref{eq-sbaradd}), (\ref{Srelprimezero}), (\ref{eq-tivadd}), and (\ref{eq-tuvadd}).

Consider a cut $V_0(y)$ of the Rindler horizon $u = 0$ and let $\mathcal{A}_{V_0}, \mathcal{A}'_{V_0}$ be the algebra of operators associated to the region $\{u = 0, v > V_0(y)\}$ and its complement respectively. Given a global state $|\psi \rangle$ we can consider its restriction to $\mathcal{A}_{V_0}$. One can then purify this restriction in different ways, including the trivial purification. We will be interested in the purification introduced in \cite{Ceyhan:2018zfg}, which is based on modular flow.

For the global vacuum $|\Omega\rangle$ recall that the full modular Hamiltonian associated to the cut $V_0$ defines a modular operator via $K_{V_0} = -\log \Delta_{\Omega; \mathcal{A}_{V_0}}$ and that $\Delta^{is}_{\Omega;\mathcal{A}_{V_0}}$ simply acts as the boost that fixes $V_0$. We note that $\Delta_{\Omega;\mathcal{A}_{V_0}}$ is related to the reduced density matrix in Eq. \eqref{Modham} by $\Delta_{\Omega; \mathcal{A}_{V_0}} = \log \sigma_{>V_0}\otimes \mathds{1}_{<V_0} - \mathds{1}_{>V_0}\otimes \log \sigma_{<V_0}$. 

For a general state $|\psi\rangle$ that is cyclic and separating, one can define the relative modular operator as \cite{Witten:2018lha,Bisognano:1975ih,Araki:1976zv}
\begin{align}
\Delta_{\psi|\Omega; \mathcal{A}_{V_0}} = S^{\dagger}_{\psi|\Omega; \mathcal{A}_{V_0}}S_{\psi|\Omega; \mathcal{A}_{V_0}}~,
\end{align}     
where 
\begin{align}
S_{\psi|\Omega; \mathcal{A}_{V_0}}\alpha |\psi\rangle = \alpha^{\dagger}|\Omega\rangle, \ \forall \alpha \in \mathcal{A}_{V_0}
\end{align}
defines the Tomita operator. 

We then purify $|\psi\rangle$ restricted to $\mathcal{A}_{V_0}$ using the Connes cocycle 
\begin{align}
|\psi_s\rangle = u'_s|\psi\rangle, \ u'_s = (\Delta'_{\Omega})^{is}(\Delta'_{\Omega|\psi})^{-is} \in \mathcal{A}'_{V_0}~.
\end{align}
The Connes cocycle can roughly be thought of as a half-sided boost that fixes the state restricted to $\mathcal{A}_{V_0}$ but stretches all of the excited modes in the complement region. Specifically, expectation values of operators in $\mathcal{A}_{V_0}$ are left invariant whereas expectation values of operators in $\mathcal{A}'_{V_0}$ are equivalent to those evaluated in the state $\Delta^{-is}_{\Omega}|\psi\rangle$. This follows (restricting to cyclic and separating states for simplicity) from the relation $(\Delta')^{is}_{\psi|\Omega}\Delta^{-is}_{\Omega|\psi} = 1$, which implies 
\begin{align}
|\psi_s\rangle = \Delta^{-is}_{\Omega}u_s|\psi\rangle~.
\end{align}
If we consider an operator $\mathcal{O}'\in \mathcal{A}'_{V_0}$ then $[u_s, \mathcal{O}'] = 0$ so 
\begin{align}
\langle \psi_s|\mathcal{O}'|\psi_s\rangle = \langle \psi| \Delta^{is}_{\Omega}\mathcal{O}'\Delta^{-is}_{\Omega}|\psi\rangle~.
\end{align}
Note that $v = V_0(y)$ is a fixed point of the boost. 

In the limit $s\rightarrow \infty$ all of these excitations become soft. More specifically, 
\begin{align}
\langle T_{vv} \rangle_s\lvert_{v < V_0(y)} &\equiv \langle \psi_s|T_{vv}(v)|\psi_s\rangle\lvert_{v <V_0(y)} \nonumber
\\ &= e^{-4\pi s}\langle \psi|T_{vv}(V_0 +e^{-2\pi s}(v-V_0))|\psi\rangle\lvert_{v < V_0(y)}
\end{align}
which just follows from the usual algebra of half-sided modular inclusions. Hence $\langle T_{vv} \rangle_s \rightarrow 0$ as $s\rightarrow \infty$ for $v < V_0(y)$.  

Not only that but also 
\begin{align}
\lim_{s\rightarrow \infty}\int_{-\infty}^v dv \ \langle T_{vv}\rangle_s \rightarrow 0, \  v < V_0(y)\label{IntegratedEnergy}~.
\end{align}

To see what this implies about the energy of the boosted side, we make use of the sum rule derived in \cite{Ceyhan:2018zfg} for null derivatives of the relative entropy: 
\begin{align}
2\pi \left(P_s - e^{-2\pi s}P\right) = \left(e^{-2\pi s}-1\right)\frac{\delta S_{\text{rel}}(\psi|\Omega; \mathcal{A}_{V})}{\delta V}\Big\lvert_{V_0} \label{sum rule}~,
\end{align}
where 
\begin{align}
P = \int_{-\infty}^{\infty} dv\  \langle T_{vv}\rangle_{\psi}
\end{align}
is the average null energy of the original state, $P_s$ is the average null energy of $|\psi_s\rangle$, and 
\begin{align}
S_{\text{rel}}(\psi|\Omega; \mathcal{A}_{V}) = -\langle \psi|\log \Delta_{\psi|\Omega; \mathcal{A}_{V}}|\psi\rangle
\end{align}
is the relative entropy of the original state for some general cut $V(y)$. 

The relative entropy can also be written as
\begin{align}
S_{\text{rel}}(\psi|\Omega; \mathcal{A}_{V}) = \langle K_{V} \rangle_{\psi}- S(V)
\end{align}
and moreover \cite{Ceyhan:2018zfg}
\begin{align}
\frac{\delta \langle K_{V}\rangle_{\psi} }{\delta V}\Big\lvert_{V_0} = -2\pi \int_{V_0(y)}^{\infty}dv \ \langle T_{vv} \rangle_{\psi}~.
\end{align}
Thus in the limit $s\rightarrow \infty$ we find, using Eq.~\eqref{IntegratedEnergy}, 
\begin{align}
\langle T_{vv}\rangle\lvert_{v \leq V_0(y)} = -\frac{1}{2\pi}\frac{ \delta S}{\delta V}\Big\lvert_{V_0}\ \delta(v - V_0(y))\end{align}
as desired. This reproduces both Eq. \eqref{eq-tvvkh} and Eq. \eqref{eq-sprimekh}. 

As a final point, note that under the Connes cocycle we also have the following properties: 
\begin{align}
\langle T_{uv}\rangle_{s\rightarrow \infty} &= \langle T_{uv}(V_0)\rangle_{\psi}~, \\ 
 \langle T_{iv}\rangle_{s\rightarrow \infty} &\rightarrow 0~.
\end{align}
This very easily reproduces the properties Eq. \eqref{eq-tivadd} and Eq. \eqref{eq-tuvadd}. 


\section{Discussion}
\label{sec-discussion}
We end by discussing the boundary interpretation of the generalized entropy of a QMT surface. We will also briefly describe future work on a systematic algorithm for constructing the states we conjectured in Sec.~\ref{sec-calculation}.
\subsection{Boundary dual}
\begin{figure}[]
\includegraphics[width=.35\textwidth]{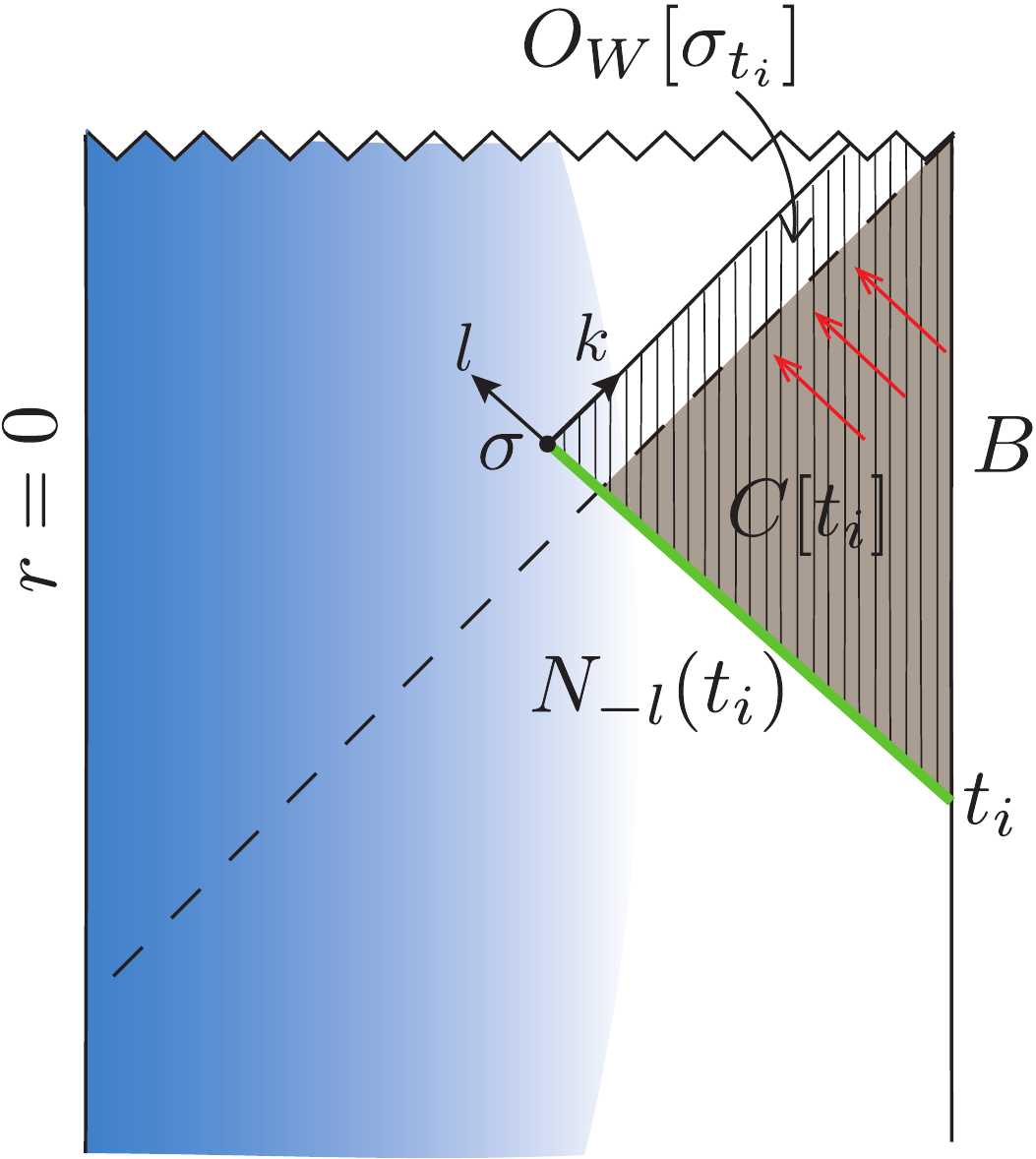}
\caption{We would like to fix the data on $N_{-l}(t_{i})$ (green thick line), while coarse-graining in the interior of the QMT surface. Simple data in the boundary region $t>t_{i}$ fixes the causal wedge $C[t_{i}]$ and thus fixes only a portion of $N_{-l}(t_i)$. In order to fix all of $N_{-l}(t_{i})$ one must allow for sources that remove the excitations (red arrows) that enter the black hole after $\sigma$; this can cause the causal wedge to grow to include $N_{-l}$. In the coarse-graining set $\mathcal{F}$, the simple data must agree for all allowed sources.}
\label{fig-simple}
\end{figure}

Within AdS/CFT, it is natural to ask whether the coarse-graining prescription for $S_{\rm outer}$ in Sec.~\ref{sec-calculation} has a boundary dual. In other words, there must exist a boundary state dual to the bulk coarse-grained semiclassical state of Sec.~\ref{sec-calculation}. Based on Eq.~\eqref{eq-bound}, we know that the boundary dual to this state is a mixed state that maximizes the boundary von Neumann entropy subject to fixing the semiclassical state in $O_{W}[\mu]$. Since in Sec.~\ref{sec-calculation} we only considered a case where we have reflecting boundary conditions at infinity, fixing $O_{W}[\mu]$ amounts to fixing the past boundary of $O_W[\mu]$, labelled $N_{-l}(t_{i})$ in Fig.~\ref{fig-simple}.

Therefore, the question of whether there is a natural boundary dual to our bulk coarse-graining prescription reduces to that of whether fixing the semiclassical state on $N_{-l}(t_{i})$ has a natural interpretation in the boundary. Our answer to this question is very similar to the simple entropy $S_{\text{simple}}$ prescription of \cite{Engelhardt:2017aux, Engelhardt:2018kcs}.\\

Since we would like to refer to the bulk as little as possible, we define the QMT surface $\mu$ associated to a time slice $t_{i}$ of the boundary by constructing an ingoing null surface from $t_{i}$ and marking the first QMT surface on it. In general, this surface could reach caustics before reaching $\mu$; Ref.~\cite{Engelhardt:2018kcs} deals with this technicality. Here we ignore this issue by restricting to special classes of states (e.g. perturbations to Killing horizons).

Let $\rho(t_{i})$ be the original boundary state at time $t_{i}$. We would like to construct a boundary state with maximum von Neumann entropy, which agrees with the semiclassical bulk state on $N_{-l}(t_{i})$. In order to accomplish this, we must find a boundary definition of $\mathcal{F}$, the set of density matrices dual to the semiclassical state on $N_{-l}(t_{i})$.

Let us first consider $\mathcal{F}$ to be the states that agree with $\rho(t_{i})$ on simple boundary observables $\mathcal{A}$ on $t>t_{i}$. Simple observables are defined to be boundary operators whose associated excitations propagate causally in the bulk \cite{Engelhardt:2017aux, Engelhardt:2018kcs}, so this data fixes the bulk causal wedge of $t>t_{i}$ ($C[t_{i}]$ in Fig. \ref{fig-simple}). However, $C[t_{i}] \subseteq O_{W}[\sigma_{t_{i}}]$, so in general this set $\mathcal{F}$ would not be constrained enough to fix all of the data on $N_{-l}(t_{i})$.

The discrepancy between  $C[t_{i}]$ and $O_{W}[\sigma_{t_{i}}]$ arises from matter that enters the black hole to the future of $\sigma_{t_i}$. This causes the event horizon to grow and lie properly inside of the outer wedge. To fix all of $O_{W}[\sigma_{t_{i}}]$ given $\rho(t_{i})$, one must turn on boundary sources that will absorb the future infalling excitations and achieve $C[t_{i}] = O_{W}[\sigma_{t_{i}}]$. This may seem acausal, but so is the definition of simple operator as an operator that can be represented by local boundary operators smeared over space {\em and time}.

Therefore, the coarse-graining set $\mathcal{F}$ should consist of the states such that the simple boundary observables $\mathcal{A}$ agree with those of $\rho(t_{i})$ even after both states have been subject to turning on various simple sources on the boundary:
\begin{align}
S_{\text{simple}}(t_{i}) = \max_{\tilde{\rho} \in \mathcal{F}}~S(\tilde{\rho})
\end{align}
with
\begin{align}
\mathcal{F} = \{\rho: \langle E \mathcal{A} E^{\dagger} \rangle_{\tilde{\rho}(t)} = \langle E \mathcal{A} E^{\dagger} \rangle_{\rho},~t \geq t_{i};~\forall E \}
\label{eq-simpleobs}
\end{align}
where $\mathcal{A}$ is the set of simple observables and $E$ denotes unitaries associated with turning on various simple boundary sources.

Note that $C[t_{i}] \subseteq O_{W}[\sigma_{t_{i}}]$ in all semiclassical states \cite{C:2013uza}. Therefore, subjecting the states to various simple sources is never going to make a slice larger than $N_{-l}(t_{i})$ causally accessible from the boundary. Given the state $\rho(t_{i})$, there exists a fine-tuned choice of sources that will make $C[t_{i}] = O_{W}[\sigma_{t_{i}}]$. But since this choice is state-dependent and difficult to specify from a pure boundary perspective, we choose the boundary coarse-graining family $\mathcal{F}$ to agree with $\rho(t_{i})$ on simple data subject to \textit{all} simple sources turned on.

So far we have defined $\mathcal{A}$ as the set of boundary observers that correspond to bulk excitations that propagate causally.  The classical analysis of Refs.~\cite{Engelhardt:2017aux, Engelhardt:2018kcs} further specified $\mathcal{A}$ to consist only of one point functions of all local operators on the boundary. This will fix the states of the classical fields in the bulk that are causally determined by the boundary region $t \geq t_{i}$. Since here we are interested in fixing the quantum state of the bulk fields on $N_{-l}(t_{i})$, our set $\mathcal{A}$ needs to include higher point function of local bulk operators.

However, we are still interested in maintaining locality in the bulk and therefore want to disallow a large density of local probes in any bulk region. This is following the expectation that such excitations would cause large backreaction and therefore a breakdown of locality~\cite{Almheiri:2014lwa}. From a boundary perspective, a local bulk operator in the causal wedge is dual to a smeared boundary operator~\cite{Hamilton:2006az}. Therefore, our set $\mathcal{A}$ needs to include all products of smeared boundary operator as long as there is not an $\mathcal{O}(N)$ number of overlap in the support of the smeared operators. This choice of $\mathcal{A}$ in Eq.~\eqref{eq-simpleobs} is a natural candidate for fixing the quantum state on $N_{-l}(t_{i})$; we leave a thorough investigation of this issue to future work.

We refer the reader to \cite{Engelhardt:2018kcs} for a careful demonstration of $S_{\text{simple}} = S_{\text{outer}}$ in the bulk classical limit. 

\subsection{Semiclassical Stretched States}
In this paper, we started from a classical construction in general relativity, whose quantum interpretation is the coarse-graining of a quantum state so that its entropy matches the area of a marginally trapped surface. We elevated this to a semi-classical conjecture that we interpret as a coarse-graining that will match the generalized entropy of a quantum marginally trapped surface, while holding fixed the exterior quantum state. In the QFT limit, our conjecture is confirmed by the limit of the CF sequence of states~\cite{Ceyhan:2018zfg}.

Thus, we were able to derive a nontrivial, testable property of QFT from a hypothetical assumption about quantum gravity. This is similar to how the QNEC was derived from the QFC, a hypothetical extension of the classical focussing property of general relativity. This is a satisfying connection. QFT has not been directly probed in this limit, and a direct verification of the CF limit or of the QNEC would constitute a test of our ideas about quantum gravity.

Interestingly, there appears to be a larger set of relations of the type we explored here. Our starting point, the EW construction, is essentially unique. However, the CF construction produces a one-parameter family of states, given an input state and a cut on a Killing horizon. Here we only made use of the limit approached by these states as the flow parameter diverges. But we expect that there exists a classical construction (which may limit to the EW construction) that matches the entire one-parameter CF family.

In the special case where the cut is a bifurcation surface of the Killing horizon, the CF construction admits an interesting intuitive interpretation: all correlators of operators restricted to the left (or to the right) behave as if we had boosted the state on the left side of the cut (but not on the right). In QFT, a one-sided boost would result in a divergent-energy shock at the cut, because it would destroy the vacuum. But the CF flow is more subtle; in a sense it boosts only the ``excited part'' of the state on $N_k^-$, while leaving the vacuum entanglement across the cut intact.

This suggests a simple classical analogue of CF flow. At the classical level, a half-sided boost is innocuous. It can be applied to initial data on the null surface $N_k^-$ with no ill consequences at the cut. 
However, a generic cut of a Killing horizon is not a bifurcation surface and hence is not a fixed point of the Killing flow.

Nonetheless, one can construct a sequence of geometries by a construction we will call the {\em left stretch}. Given the state and affine parameter $v$ on the entire Killing horizon $N_k$, rescale $V \to V'=e^{s}V$ on the left side $N_k^-$, and do nothing on the right: $V\to V'$ on $N_k^+$.  This will rescale all $v$-derivatives of classical fields by $e^{-s}$. To preserve the inner product $k^a l_a\equiv g_{ab}(\partial_v)^a(\partial_u)^b=-1$, rescale the $u$-derivatives at constant $(v,y)$ by $e^s$. Then glue the two halves back together, treating $V'$ as a true affine parameter.

For the full initial data on $N$, we need to know not only the intrinsic geometry but also $\theta_l$, the expansion in the null direction off of $N_k$. This is obtained by holding $\theta_l$ fixed on $N_k^+$ and integrating the cross-focussing equation,
\begin{align}
  k^a\nabla_a \theta_l = -\frac{1}{2} {\cal R}  - \theta_k\theta_l +\chi^2+\nabla\cdot\chi
  + 8\pi G\, T_{kl}~,
\end{align}
to obtain $\theta_l$ on $N_k^-$. Since all terms on the right hand side scale trivially, this rescales the difference $\theta_l-\theta_l|_{V_0}$ by $e^s$.

Because $\theta_l$ is not given by a simple rescaling unless $\theta_l|_{V_0}=0$, the left stretch results in physically inequivalent initial data {\em even in the left exterior of $\sigma$ alone}. The intrinsic data on $N_k^-$ are stretched, as measured by a ruler defined by the evolution of the extrinsic curvature $\theta_l$. 

Interestingly, the left stretch is physically sensible if and only if the cut is a trapped surface. This is because the expansion $\theta_k$ along $N_k$ is determined not only by the left stretch itself, but also by the Raychaudhuri equation, and the two methods must agree.
Let the inaffinity $\kappa$ be defined by  $k^b \nabla_b k^a = \kappa k^a$. Affine parametrization corresponds to $\kappa = 0$ everywhere. The left stretch implements
\begin{equation}
  V(y) \rightarrow e^{sH[-V(y)+V_0(y)]} V(y)~,
\end{equation}
where $H(v)$ is the Heaviside step function and $v=V_0(y)$ is the marginally trapped surface $\sigma$. This generates a non-zero inaffinity
\begin{equation}
  \kappa = (1-e^{-s})\delta[V(y)]~.
  \label{eq-inaf}
\end{equation}
The Raychaudhuri equation for non-affine parametrization reads
\begin{align}
k^a \nabla_a \theta_k = -\frac{1}{2}\theta_k^2 - \varsigma_k^2 -\kappa \theta_k -8\pi G \ T_{kk}~.
\end{align}
We insist that the new parameter $V'$ be treated as affine, which means we are demanding that the inaffinity term $\kappa\theta$ vanishes even after the left stretch. 
By Eq.~(\ref{eq-inaf}), this will be the case if and only if $\theta_k=0$ at the cut.

Importantly, Eqs.~(\ref{eq-tkis0}),  (\ref{eq-choices1}) and (\ref{eq-choices2}) become satisfied in the limit as $s\to\infty$. These are precisely the conditions imposed by EW for the classical coarse-graining construction. In this sense the left stretch can be viewed as generating a one-parameter interpolation from the original initial data to the coarse-grained data.\footnote{However, there are interesting differences to the EW analysis. For example, the left stretch yields divergent $T_{uu}$, as does the CF limit. Yet, EW argue that this can be avoided. There may be a larger family of relevant states.}

We close with two brief remarks. At the level of semiclassical gravity, the left stretch should naturally combine with the CF construction, so that not only the geometric and classical data, but also the quantum initial data are stretched. Moreover, we expect that the left stretch (applied classically to the RT or semiclassically to the quantum RT surface) is the gravity dual of the CF flow applied to the boundary of Anti-de Sitter space.\\

\acknowledgments We thank Netta Engelhardt, Thomas Faulkner, and Aron Wall for discussions. This work was supported in part by the Berkeley Center for Theoretical Physics; by the Department of Energy, Office of Science, Office of High Energy Physics under QuantISED Award DE-SC0019380 and contract DE-AC02-05CH11231; and by the National Science Foundation under grant PHY-1820912.

\onecolumngrid

\begin{figure}[]
\includegraphics[width=.8\textwidth]{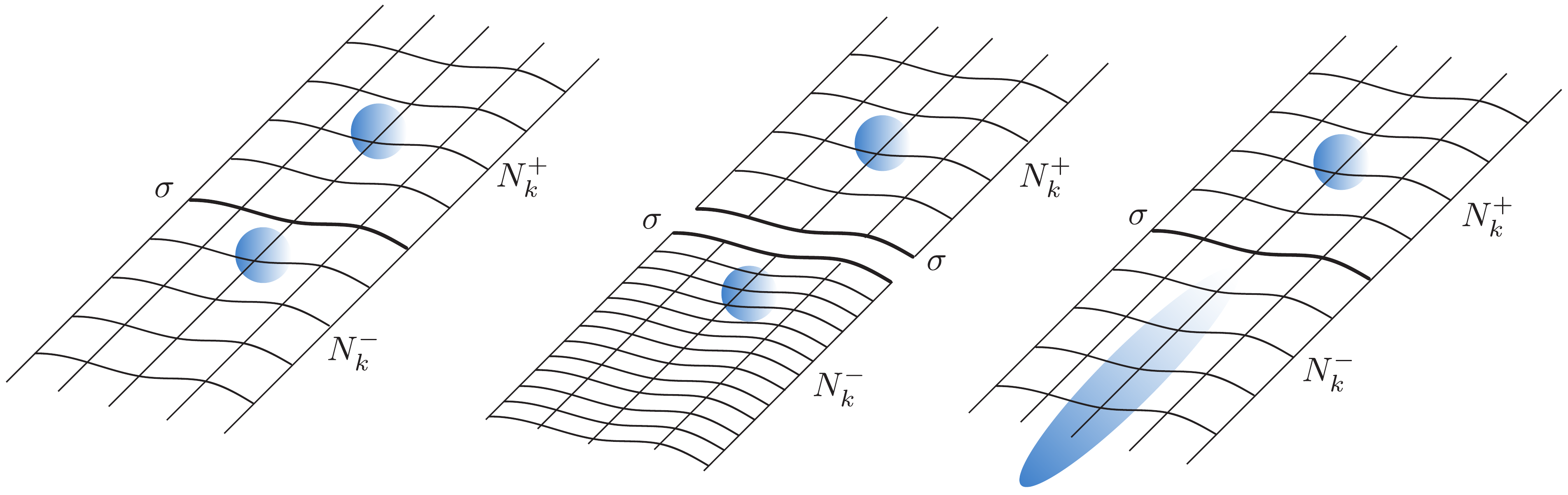}
\caption{The left stretch is a classical analogue of the CF flow that generalizes it to nontrivial geometries. Left: The null surface $N_k$ split by the marginally trapped surface $\sigma$. Middle: The affine parameter is rescaled on $N_k^-$ but held fixed on $N^+_k$. This is the same initial data in nonaffine parametrization. Right: The two pieces are glued back together, treating the new parameter as affine. This yields inequivalent initial data.\label{fig-leftstretch}}
\end{figure}

\twocolumngrid

\appendix
\section{Ant Conjecture and Properties of Energy Minimizing States}
\label{sec:main}


In Sec.~\ref{sec-cf}, we showed that the Ceyhan-Faulkner construction proves our conjecture in the pure-QFT limit. The original purpose of the CF construction, however, was to prove Wall's ``ant conjecture''~\cite{Wall:2017aa} (and thus, the Quantum Null Energy Condition~\cite{Bousso:2016aa}). 
It is therefore of interest to ask how closely related our coarse-graining conjecture is to the ant conjecture on Killing horizons.
It is easy to see that Eqs.~(\ref{eq-tvvkh}) and (\ref{eq-sprimekh}) imply the ant conjecture. Conversely, we will show in this section that the ant conjecture implies Eqs.~(\ref{eq-tvvkh}) and (\ref{eq-sprimekh}), but only in 1+1 dimensions.

In Appendix~\ref{sec-ant}, we will review the ant conjecture. In Appendices~\ref{sec-prop} and \ref{sec-high}, we establish some general properties that energy-minimizing states must satisfy. We show that the minimum energy completion has vanishing stress tensor on the unconstrained half-space, with all of the remaining energy appearing as a shock immediately on the cut. We also show that for a pure minimum energy state, the von Neumann entropy of semi-infinite regions is constant so long as the region's boundary lies on the unconstrained side. In 1+1 dimensions, we can also show that the integrated left stress tensor vanishes. Thus the ant conjecture implies Eqs.~(\ref{eq-tvvkh}) and (\ref{eq-sprimekh}), the key properties of the field theory limit of our coarse-graining conjecture. In higher dimensions, we are unable to establish this result.

\subsection{Ant Conjecture}
\label{sec-ant}

\begin{figure}[]
\includegraphics[width=.45\textwidth]{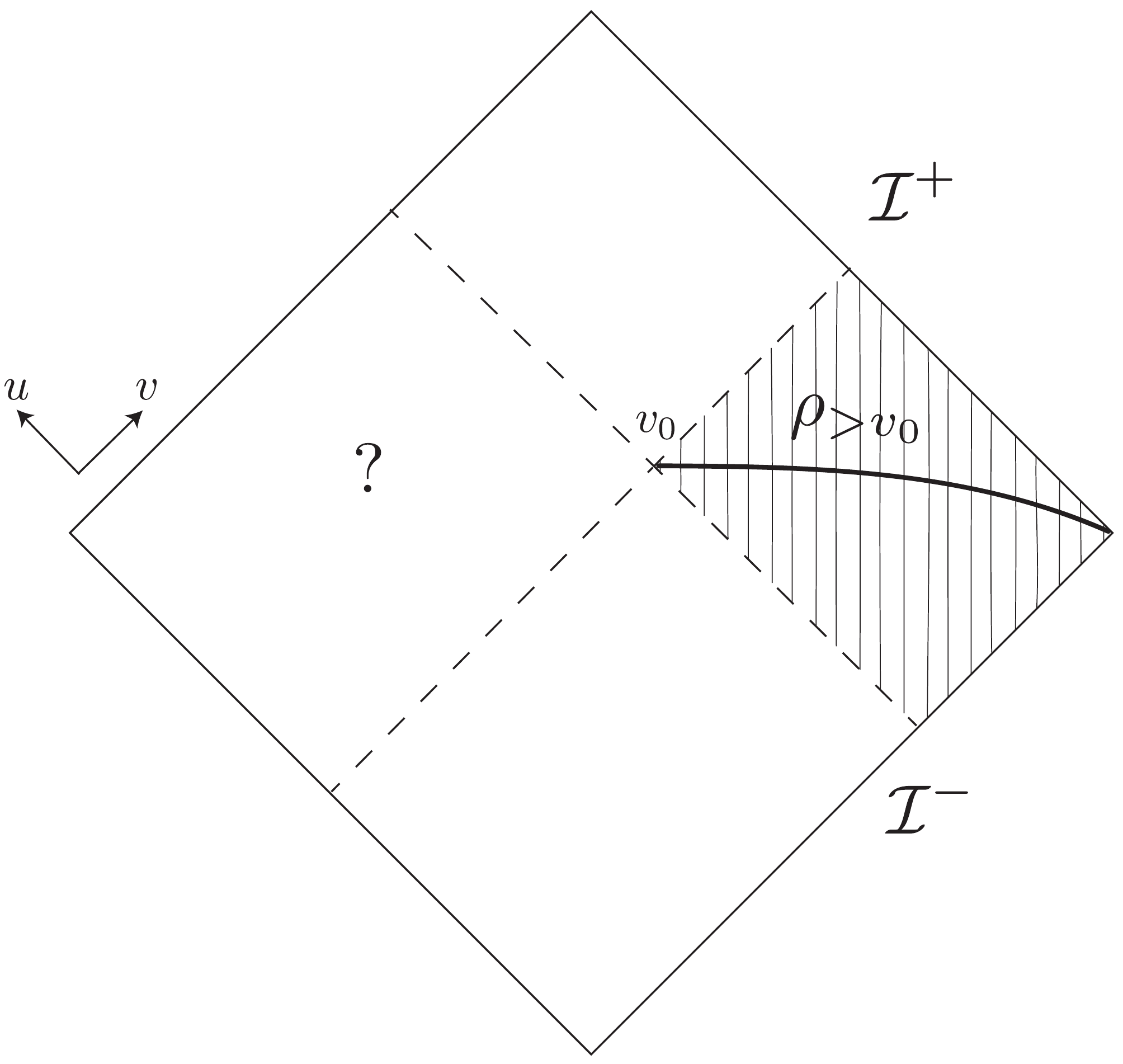}
\caption{The ant conjecture in 1+1 dimensions. A left-walking ant has access to all the information in the right wedge. It asks what is the least amount of additional energy it might still encounter to the left of $v_0$. The conjecture states that this is $\hbar S'/2\pi$, where $S'$ is the right derivative of the von Neumann entropy of the reduced state on the right, evaluated at the cut. We show that this statement is equivalent to the nongravitational limit of our coarse-graining conjecture. \label{fig-2dant}}
\end{figure}

Wall's ``ant argument'' for the Quantum Null Energy Condition in 1+1 dimensions invokes an ant that has walked left from $+\infty$ to $v_0$. (See Fig.~\ref{fig-2dant}.) That is, given a global state $\rho$, the ant has knowledge only of the right half-space state $\rho_{>v_0}$. Pausing for rest, the ant contemplates how much energy it might still encounter in the remainder of its path, the interval $(-\infty, v_0]$. Because of global energy conditions, this amount is bounded from below.

Let $M(v_{0})$ be the lowest energy of any global state that reduces to the same $\rho_{>v_0}$.\footnote{We should point out two differences in our conventions compared to~\cite{Wall:2017aa}. First, we have switched the side on which the state is held fixed, from left to right. Secondly, in~\cite{Wall:2017aa}, $M$ was the infimum of the energy density integrated only over the complement of that fixed half-space, whereas here it is the infimum the global energy. This choice is more convenient as otherwise the presence of distributional sources at the cut $v_0$ would lead to ambiguities and require a more elaborate definition. In this respect, our conventions agree with~\cite{Ceyhan:2018zfg}.} More precisely,
\begin{align}\label{defM}
  M(v_0) \equiv \text{inf}_{\hat\rho} \left[
  \int_{-\infty}^\infty d\tilde v \, \langle T_{vv}\rangle \bigr\rvert_{\hat\rho}\right]~.
\end{align}
The infimum is over all global states $\hat\rho$ that agree with $\rho$ in the region $v>v_0$: $\mathrm{Tr}_{\leq v_0}\, \hat\rho=\rho_{>v_0}$. A strictly larger set of global states will agree with $\rho$ on a smaller region, $\rho_{v_1}$, $v_1>v_0$, so the infimum can only decrease with $v$:
\begin{equation}
  \partial_v M(v) \leq 0~.
  \label{eq-mmono}
\end{equation}

One can readily establish a lower bound on $M(v)$. The global energy appearing in the infimum can be written as $(\hbar/2\pi)(\partial_v\bar K-\partial_v K)$, by Eq.~(\ref{eq-pvk}) and its left analogue.  Moreover, Eq.~(\ref{eq-kss}) must hold for all states appearing in the infimum, so by adding $\partial_v K$ to it one finds that
\begin{equation}
  M(v_0) \geq - \frac{\hbar}{2\pi}\partial_v S_\text{rel}|_{v_0}~,
  \label{eq-mgeq}
\end{equation}
where we have used Eq.~(\ref{eq-srelks}). Note that the lower bound is determined solely by the input state $\rho$.

Wall conjectured~\cite{Wall:2017aa} that this inequality is saturated:
\begin{align}\label{conj}
  M(v_0) = -\frac{\hbar}{2\pi} \partial_{v} S_\text{rel}|_{v_0} ~.
\end{align}
This conjecture is equivalent to the existence of a sequence of states $\hat\rho^{(n)}$, all of which reduce to $\rho_{>v_0}$ on the right, such that
\begin{equation}
  \lim_{n\to\infty}\int_{-\infty}^\infty d\tilde v \langle T_{vv}\rangle|_{\hat\rho^{(n)}} =
  -\frac{\hbar}{2\pi} \partial_{v} S_\text{rel}|_{v_0} ~.
\label{eq-seq}
\end{equation}
Here we will assume the conjecture to be true. We will be interested in certain universal properties of the states in this sequence that emerge in the limit as $n\to \infty$.

\subsection{Properties of the minimum energy completion in 1+1 dimensions}
\label{sec-prop}

For compactness of notation, we will ascribe any limiting properties of the states $\hat\rho^{(n)}$ as $n\to \infty$ to a ``limit state'' $\hat\rho^\infty$. We stress that such a state need not exist. Rather, $\hat\rho^\infty$ is shorthand for $\lim_{n\to\infty}\hat\rho^{(n)}$, where the limit should be moved outside of any maps of the state to other quantitities. Moreover, we indicate $\hat\rho^\infty$ as the argument of a map by the superscript $\infty$. For example,
\begin{equation}
  S^\infty(v_0) = \lim_{n\to\infty}\left[ - \text{Tr}\, \hat\rho^{(n)}_{>v_0} \log \hat\rho^{(n)}_{>v_0}\right]~.
\end{equation}

By Eq.~(\ref{conj}) and the discussion leading to Eq.~(\ref{eq-mgeq}), the state $\hat\rho^\infty$ must saturate both inequalities in Eq.~(\ref{eq-kss}):
\begin{equation}
  \partial_v \bar K^\infty |_{v_0}  = \partial_{v}\bar S^\infty |_{v_0} = \partial_{v}S^\infty |_{v_0} ~.
\end{equation}
The first equality implies
\begin{equation}
  \partial_v \bar S_\text{rel}^\infty |_{v_0}=0~.
\end{equation}
Applying the left analogues of Eqs.~(\ref{eq-mmono}) and (\ref{conj}) to $\bar M$ (with $\hat\rho^\infty$ as the input state!), we have
\begin{equation}
  \partial_v^2 \bar S_\text{rel}^\infty \geq 0~,
\end{equation}
for all $v$. The above two consequences of Wall's conjecture, combined with positivity and monotonicity of the left relative entropy,
\begin{eqnarray} 
  \bar S_\text{rel}^\infty & \geq & 0~,\\
  \partial_v \bar S_\text{rel}^\infty & \geq & 0~,
\end{eqnarray}
imply that
\begin{align}\label{important}
  \partial_{v} \bar S_\text{rel}^\infty =0 \text{~for all~} v<v_{0}~.
\end{align}
This is a very strong condition and it intuitively suggests that for $v<v_{0}$ we have a vacuum-like state. In particular all local observable in the region between $v$ and $v_{0}$ for $v<v_{0}$ need to register vacuum values otherwise we would have $S_{\text{rel}}^{\infty}(v_{0}) > S_{\text{rel}}^{\infty}(v)$. This is particular tells us that 
\begin{align}
\langle T_{vv}(v)\rangle|_{\hat\rho^{\infty}}=0 \text{ for } v< v_0~.
\end{align}
The above equation combined with Eq. \eqref{important} implies
\begin{align}
\partial_{v}^2 \bar{S}^{\infty} = 0 \implies \partial_{v} \bar{S}^{\infty} = \alpha \text{ for } v\leq v_0~,
\end{align}
In fact, in 1+1 CFTs we can argue that $\alpha=0$ by invoking the strengthened version of the QNEC \cite{Wall:2011kb, Koeller:2016aa} \footnote{We thank Aron Wall for suggesting the use of strengthened QNEC here}:
\begin{align}\label{strongqnec}
\langle T_{vv} \rangle \geq \frac{\hbar}{2\pi}\partial_{v}^{2} \bar{S} + \frac{6\hbar}{c} (\partial_{v}\bar{S})^{2}~.
\end{align}
Now, Eq.~\eqref{important} implies that 
\begin{align}\label{intermediate}
\langle T_{vv} \rangle = \frac{\hbar}{2\pi}\partial_{v}^{2} \bar{S} \text{    for $v<v_{0}$}~,
\end{align}
which together with Eq.~\eqref{strongqnec} implies that $\partial_{v}\bar{S} = 0$. So, we conclude that for $v<v_{0}$,
\begin{align}\label{zeroT2d}
  \partial_{v}\bar{S}^{\infty} = 0 \text{  and  } \partial_{v} \bar{S}_{\text{rel}}^{\infty}=0 & \implies \\
  & \lim_{\epsilon\to0}\int_{-\infty}^{v_{0}-\epsilon} d\tilde{v} \, \langle T_{vv}\rangle \bigr\rvert_{\hat{\rho}^{(\infty)}} = 0~.
\end{align}
We also know that
\begin{align}
  \lim_{\epsilon\to 0} \left[
  \int_{-\infty}^{v_{0}+\epsilon} d\tilde{v} \, \langle T_{vv}\rangle \bigr\rvert_{\hat{\rho}^{\infty}}\right] = \frac{\hbar}{2\pi} \partial_{v} S\bigr\rvert_{v_{0}}~.
\end{align}
This along with Eq.~\eqref{zeroT2d} implies that the minimum energy state contains a shock (a delta function in energy density) at $v_{0}$, and vanishing energy to its left:
\begin{align}
\langle T_{vv} \rangle = \left(\frac{\hbar}{2\pi}
  \partial_{v}S\bigr\rvert_{v_{0}}\right) \delta(v-v_{0}) \text{  for
  } v\leq v_{0}~.
  \label{eq-tdelta}
\end{align}
If $\hat\rho^\infty$ is a pure state\footnote{The conclusion would extend to mixed states under the assumption that $\Delta S(v)$ remains bounded from below for any $v$ in the limit as $n\to\infty$. The status of this assumption is not clear to us, however.} this further implies that
\begin{equation}
  \partial_v S=0 \text{ for } v< v_0~.
  \label{eq-sconst}
\end{equation}
In fact, we expect that $\hat\rho^\infty$ can always taken to be pure. The basic idea is that any density operator can be purified by a suitable auxiliary system. In general the auxiliary system has to be external, but we now argue it can be taken to be distant soft modes in the quantum field itself.

Suppose we had identified a sequence $\hat\rho^{(n)}$ that limits to a mixed $\hat\rho^\infty$. Finiteness of the energy requires that each state in the sequence looks like the vacuum in some sufficiently distant left region $v<v^{(n)}$ with $v^{(n)}<v_0$. We can take $v^{(n)} \to -\infty$ as $n\to\infty$. We can add a purification of the state $\hat\rho^{(n)}$ in soft wavepackets localized to the region $v<v^{(n)}$. This results in a new, pure state and we redefine $\hat\rho^{(n)}$ to be that state. Since we have not modified the state in the region $v>v_0$, it will still reduce to the given right state $\rho_{>v_0}$; and since the region $v<v^{(n)}$ is semi-infinite, we can take the purifying wave-packets to have arbitrarily small energy. In particular, we can take their contribution to the energy to vanish in the limit as $n\to\infty$.

\subsection{Higher-dimensional case}
\label{sec-high}

\begin{figure}[]
	\includegraphics[width=.45\textwidth]{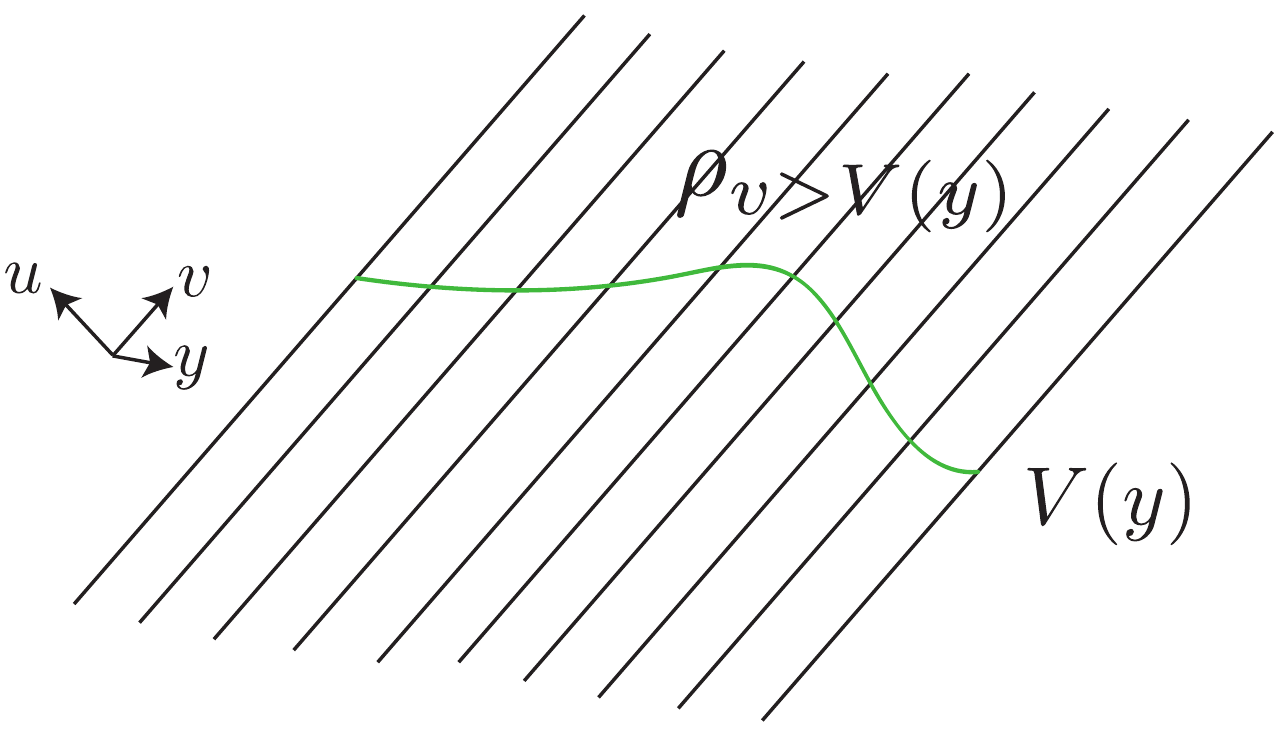}
	\caption{A general cut of the Rindler horizon in $d > 2$. An army of ants marches down along the null direction towards the cut. Given the state above the cut, they ask what is the minimum energy still to come. \label{fig-higherdimregion}}
\end{figure}
The generalization of the above result to higher dimensions is straightforward. We can consider any Killing horizon $N=\mathbf{R}\times {\cal B}$, with $v\in \mathbf{R}$ an affine parameter along light-rays orthogonal to the $d-2$ dimensional spatial surface ${\cal B}$ with collective coordinates $y$.

The analogue of the 1+1 dimensional ant is now an army of ants that have walked along the null generators from $v=+\infty$ to the position $v=V(y)$, so that they know the state $\rho_{>V(y)}$. (See Fig.~\ref{fig-higherdimregion}.) The ants again ask about the minimum global energy consistent with this knowledge, $M[V(y)]$. This quantity can only decrease under deformations of $V(y)$ that are everywhere positive:
\begin{equation}
  \frac{\delta M}{\delta V(y)}\leq 0~.
\end{equation}

The definition of $M$ differs from the 1+1 case only through an additional transverse integral over $d^{d-2}y$. It can be shown~\cite{Koeller:2017aa,Casini:2017aa} that the modular Hamiltonian, too, is simply the sum of the local Rindler energies associated with the individual null generators, Eq.~(\ref{eq-k11}):
\begin{equation}
  \Delta K (V_{0}(y))  = \frac{2\pi}{\hbar}\int d^{d-2}y \int_{V_0(y)}^\infty dv\, (v-V_0(y)) T_{vv}~,
\end{equation}
By the analogue of Eq.~(\ref{eq-kss}),
\begin{align}
\frac{\delta \bar{K}}{\delta V(y)} \geq \frac{\delta \bar{S}}{\delta V(y)} \geq \frac{\delta S}{\delta V(y)}~,
\end{align}
one finds
\begin{equation}
  M \geq -\frac{\hbar}{2\pi}\frac{\delta S_\text{rel}}{\delta V(y)}~.
  \label{eq-mhd}
\end{equation}

The ant conjecture again demands that this be an equality. That is, there exists a global state $\hat{\rho}^{\infty}$ that saturates Eq.~(\ref{eq-mhd}) (or if not, saturation can at least be approached, in the limit of a sequence of global states). The same arguments as in the 1+1 dimensional case imply that $\hat{\rho}^{\infty}$ satisfies
\begin{align}\label{dimportant}
\frac{\delta \bar{S}^{\infty}_{\text{rel}}}{\delta V(y)}=0~ \text{for all}~ v<V_{0}(y)~.
\end{align}
Exactly as in the 1+1 case, the above condition implies
\begin{align}
  \langle T_{vv}(v)\rangle|_{\hat\rho^{\infty}} &=0 ~~\text{ for }~ v<V_{0}(y)~,
  \label{eq-jkj} \\
\frac{\delta^{2}\bar{S}^{\infty}}{\delta V(y_{1}) \delta V(y_{2})} &= 0 \implies \frac{\delta \bar{S}^{\infty}}{\delta V(y)} = \alpha ~~\text{for all}~ v<V_{0}(y)~.
\end{align}
where $\alpha$ is some constant. As was discussed at the end of the previous section, we can take $\rho^{\infty}$ to be a limit of pure states where we additionally have 
\begin{align}
\frac{\delta S^{\infty}}{\delta V(y)} = \alpha ~~\text{for all}~ v<V_{0}(y)~,
\end{align}

At this point, it would be nice to argue that $\alpha=0$ as in the 1+1 dimensional case, but we will leave this to future work. If we assume that $\alpha=0$, then the purity of the global state implies
\begin{align}
\frac{\delta \bar{S}^{\infty}}{\delta V(y)} = 0 ~~\text{for all}~ v<V_{0}(y)~,
\end{align}
and together with Eq. \eqref{dimportant} one obtains
\begin{align}
  \lim_{\epsilon\to 0} \left[
  \int_{-\infty}^{V_{0}(y)-\epsilon} d\tilde{v} \, \langle T_{vv}\rangle \bigr\rvert_{\hat{\rho}^{\infty}}\right]  = 0 ~,
  \label{eq-klj}
\end{align}
for all $y$.
Note that Eq.~(\ref{eq-klj}) does not otherwise follow from Eq.~(\ref{eq-jkj}): because $\hat\rho^\infty$ is defined as a limit of a sequence, it would be possible for $\langle T_{vv}\rangle$ to approach zero while its integral approaches a finite value. Assuming the ant conjecture, that Eq.~(\ref{eq-mhd}) is an equality, it follows that
\begin{eqnarray} 
  \label{zeroTd}
  \langle T_{vv}(v,y) \rangle\bigr\rvert_{\hat{\rho}^{\infty}} & = & \left(\frac{\hbar}{2\pi} \frac{\delta S}{\delta V(y)} \bigr\rvert_{V_{0}}\right) \delta(v-V_{0}(y)) \nonumber \\ & & \text{for}~ v\leq V_{0}(y)~.
\end{eqnarray}

To summarize, in 1+1 dimensions, the ant conjecture implies the key properties of the coarse-graining states we conjectured: Eqs.~(\ref{eq-tvvkh}) and (\ref{eq-sprimekh}) hold on a Killing horizon. In greater than 1+1 dimensions, this implication obtains only with the unproven assumption that $\alpha=0$ above.

\bibliographystyle{utcaps}
\bibliography{all}
\end{document}